\documentclass[10pt]{article}
\usepackage[preprint]{tmlr}

\usepackage{amsmath,amsfonts,bm}

\def\eqref#1{equation~\ref{#1}}

\def\1{\bm{1}}

\DeclareMathAlphabet{\mathsfit}{\encodingdefault}{\sfdefault}{m}{sl}
\SetMathAlphabet{\mathsfit}{bold}{\encodingdefault}{\sfdefault}{bx}{n}

\usepackage{hyperref}
\usepackage{url}
\usepackage{graphicx}
\usepackage{booktabs,longtable,array,calc,tabulary,multirow}
\usepackage{amsmath,amssymb}
\usepackage{placeins}
\newcommand{\onenna}{\mbox{1-NNA}}
\def\openreview{\url{https://openreview.net/forum?id=XXXX}}
\title{Are Learned Vascular Tree Generators Better Than Conditional Resampling?}
\author{\name Reza Akbarian Bafghi\thanks{Equal contribution.} \email reza.akbarianbafghi@colorado.edu \\
      \addr University of Colorado, Boulder
      \AND
      \name Sukirt Thakur\footnotemark[1] \email sukirt@angioinsight.com \\
      \addr AngioInsight, Inc.
      \AND
      \name Maziar Raissi \email maziar.raissi@ucr.edu \\
      \addr University of California, Riverside}

\def\month{08}
\def\year{2026}
\begin{document}
\maketitle
\begin{abstract}
Fidelity scores compare generated vascular trees with held-out real trees, so by themselves they cannot detect reuse of training data. A training-free conditional resampler shows this. On coronary trees (ImageCAS) and on IntrA, about half of its branch records (branches with their parent context) are near-copies of training records. Yet on our two fidelity scores, \onenna{} on tree summaries and a parent-daughter dependence AUC, it has better point estimates than three learned baselines as retrained here. A training copy scores better still. Adding noise to the copied values traces a curve of fidelity against the measured branch-record near-copy rate. At their own coronary rates (at most 1.3\%), two learned baselines lie above this curve (worse) on parent-daughter relations. VesselGPT is inconclusive against the curve, as it is against the resampler on IntrA and on scale-free tree features. The curve is a reference, not a bound: real validation trees lie below it on both coronary scores, so low near-copy rates are not a goal in themselves. Jittered training copies also lie below it. On coronary data, one with a near-copy rate of about 1\% does so on the dependence AUC, and so does one whose stronger jitter ruins its tree summaries. On IntrA, a lightly jittered copy lies below it on \onenna{}. A matched near-copy rate therefore certifies neither score against perturbed copies. The study is exploratory. We offer the near-copy rate as a diagnostic, reported beside fidelity with three controls: a training copy, a jittered copy and real held-out trees.
\end{abstract}

\section{Introduction}\label{sec:intro}

\begin{figure}[!htb]\centering
\includegraphics[width=\linewidth]{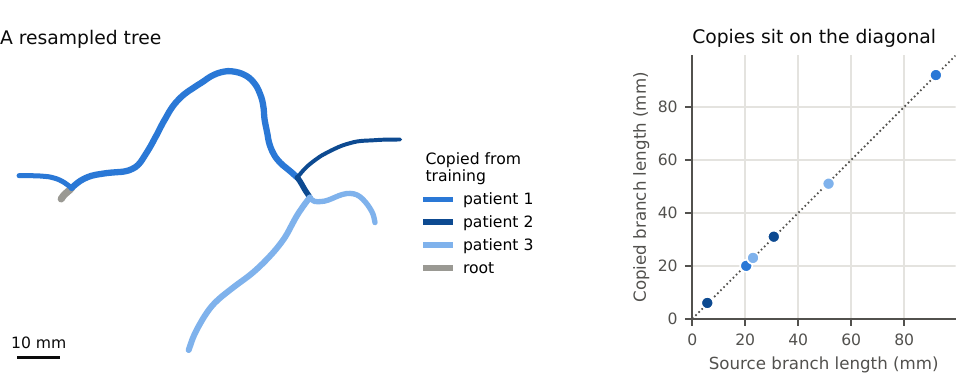}
\caption{\textbf{A resampled tree is a patchwork of training patients.} Left: a right coronary tree generated by the resampler, each daughter pair coloured by the training patient it was copied from (grey: root); 2D projection on the tree's first two principal axes. The tree (three bifurcations) is the first in sampling seed 0 with 3--4 bifurcations whose sources are all different patients. Right: the length of each copied branch against the length of its training source; every point lies on the diagonal (largest gap 0.45 mm). Radii are copied as a ratio to the parent plus a slope, so absolute radii differ from the source, and the 3D path is drawn afresh. \label{fig:1}}
\end{figure}

Figure~\ref{fig:1} shows a right coronary tree that no model learned to draw. It was grown by lookup: at each branch, we found training bifurcations whose parent branch resembled the current one and copied the two daughters of one of them. Its three daughter pairs come from three different training patients, and every copied branch length matches its source to within 0.45 mm. The tree is new; its branch descriptors are not. Yet on held-out coronary trees derived from ImageCAS \citep{zeng2023imagecas}, trees grown this way score closer to real trees than three published learned generators retrained from the authors' code. This paper asks what that result means for how tree generators are evaluated.

Synthetic vascular trees are wanted for what real cohorts lack: more anatomy. They can populate virtual cohorts for in-silico trials \citep{viceconti2021}, provide ground truth for image analysis \citep{hamarneh2010} and enlarge training sets. Learned generators now produce such trees as centerlines with radii: VesselVAE encodes a tree with a recursive variational autoencoder \citep{vesselvae2023}, VesselGPT serialises a tree into node tokens and samples them autoregressively \citep{vesselgpt}, and PartVessel generates a key graph and then its vessel segments \citep{partvessel}. Progress is judged by fidelity: a generated set is compared with held-out real trees, for example by 1-nearest-neighbour accuracy (\onenna{}; VesselGPT computes it on mesh points, we on 11 summary features per tree).

\begin{figure}[!htb]\centering
\includegraphics[width=\linewidth]{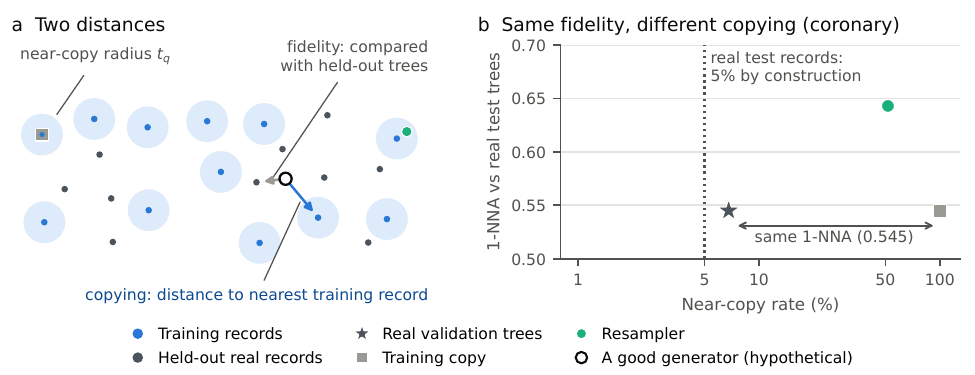}
\caption{\textbf{A training copy and real validation trees get the same \onenna{}; only the near-copy rate separates them.} (a) Schematic: discs mark the near-copy radius $t_q$ (Section~\ref{metrics}) around training records; a training copy, and a resampled record, can sit inside a disc. (b) Coronary test set: \onenna{} against the share of near-copied branch records ($q = 5$). The training copy (100\% near-copies) and real validation trees (6.8\%) have the same \onenna{}, 0.545; the resampler reaches 0.643 with 51.7\%. Dotted: real test records, 5\% by construction. \label{fig:concept}}
\end{figure}

A generator that reassembles training bifurcations may offer less diversity than its sample count suggests: recombination can add whole-tree variation while reusing local branch descriptors, and the uses above depend on variation. The sharper problem is ranking. One might answer that resembling real anatomy is the whole point, so a generator that scores well has done its job. But fidelity scores compare the generated set with held-out trees from the same population; the training set never enters the computation (Figure~\ref{fig:concept}a). Measured against held-out patients, a copied training tree is simply one more real tree, and a tree stitched from real bifurcations is made of real parts. On coronary data a verbatim training copy and real validation trees reach the same \onenna{}, although every branch record of the copy is a near-copy of a training record and only a few percent of the real trees' are (Figure~\ref{fig:concept}b). Fidelity alone therefore cannot establish novelty or freedom from training reuse, and copying can lead a leaderboard built on it.

Copying must then be measured where it happens. A resampled tree is new as a whole, so a whole-tree distance to the training set sees little; its reuse sits in single bifurcations. We therefore measure it on branch records, a branch with its parent and closest sibling, and count how many generated records are closer to a training record than nearly all real held-out records are.

We treat the procedure that grew Figure~\ref{fig:1}, a training-free conditional resampler (\emph{the resampler}), as a reference method. We put three questions to it and to VesselGPT, VesselVAE and PartVessel as retrained here (\emph{the learned baselines}), on the coronary trees and on IntrA \citep{yang2020intra}.

\textbf{Q1: who tops the fidelity leaderboard?} The resampler, although it optimises no parameters. Its lead over every learned baseline is detected on coronary data and over two of the three on IntrA, but on scale-free tree features no \onenna{} difference from VesselGPT is detected (Section~\ref{sec:q1}).

\textbf{Q2: is the winner generating something new?} Only in part: about half of its branch records are near-copies of training records, and neither fidelity score notices (Section~\ref{sec:q2}).

\textbf{Q3: at the same near-copy rate, who wins?} No learned baseline lies below the fidelity-versus-copying curve traced by the resampler with noise on coronary data: VesselVAE and PartVessel lie above it on parent-daughter relations, and VesselGPT is inconclusive. Held-out real trees and perturbed training copies do lie below it, so a matched near-copy rate informs the comparison but certifies neither score against perturbed copies (Section~\ref{sec:q3}).

The answer to Q2 is the robust core of the paper: it rests on a verbatim training copy and the resampler, neither of which is learned. Q1 and Q3 apply only to the learned baselines as trained under our budgets (Section~\ref{training-budgets-of-the-learned-models}) and to our two fidelity scores, \onenna{} on tree summaries and a parent-daughter dependence AUC; both scores read branch values of the kind the resampler copies, so its lead is partly by construction. The study is exploratory; its scope is set out in Appendix~\ref{app:disclosures}.

\textbf{Contributions.} First, a diagnostic: the branch-record near-copy rate next to every fidelity score, and the gap to the fidelity-versus-copying curve of the resampler with noise level $\lambda$, reported with three controls (training copy, jittered copy, real held-out trees; Appendix~\ref{app:disclosures}). Second, evidence on two datasets that our fidelity scores do not penalise copied branch records, which the near-copy rate detects as proximity to training records and the whole-tree data-copying test \citep{meehan2020} does not.

Section~\ref{sec:related} places the work among copying tests and example-based synthesis. Section~\ref{sec:setup} defines the resampler, the data, the controls and the scores. Sections~\ref{sec:q1} to~\ref{sec:q3} answer the three questions in turn, and Section~\ref{sec:discussion} states the limits once and turns the findings into a protocol.

\section{Related work}\label{sec:related}

Each part of our protocol has precedents; what is new is measuring copying per branch record and comparing the learned baselines with the resampler at the same measured near-copy rate.

\textbf{Copying and memorisation.} Fidelity scores can reward memorisation. For point clouds, a baseline that memorises a subset of the training data scores well on coverage and minimum matching distance \citep{achlioptas2018}, and the VesselGPT authors note that an extremely small matching distance may suggest memorisation \citep{vesselgpt}. \citet{alaa2022} observe that resampling training data can reach perfect fidelity and diversity, and score each synthetic sample for authenticity. Their authenticity score treats a generator as a mixture that either returns a training sample with noise added or produces a new one. Image diffusion models can reproduce training images \citep{carlini2023,somepalli2023}. The data-copying test $C_T$ of \citet{meehan2020}, which Section~\ref{sec:q2} applies to whole trees and to branch records, compares generated and held-out distances to the nearest training point by a per-cell Mann-Whitney statistic, negative when generated samples are closer. The near-copy rate (Section~\ref{sec:setup}) belongs to the same distance-to-training family; we propose it as a diagnostic, reported beside fidelity together with its controls, not as a better test.

\textbf{Example-based synthesis and smoothed resampling.} Texture synthesis grows an output by copying exemplar pieces that match the current context \citep{efros1999}. The nearest-neighbour bootstrap resamples the successor of a similar past state \citep{lall1996}, and the smoothed bootstrap adds noise to resampled values \citep{silverman1987}. The resampler carries both ideas to the branch records of trees: each branch receives the daughters of a similar training parent (Section~\ref{sec:setup}, Figure~\ref{fig:2}). We use it as a reference against which the learned baselines are compared, not as a proposed generator.

\textbf{Vascular tree generation.} Rule-based methods grow trees under physiological constraints \citep{karch1999,schneider2012}, and VascuSynth synthesises trees together with volumetric images \citep{hamarneh2010}. Learned generators include VesselVAE \citep{vesselvae2023}, VesselGPT \citep{vesselgpt} and PartVessel \citep{partvessel}, the three we retrain as the learned baselines (Appendix~\ref{app:baselines}). Their evaluations compare generated with real shapes: histograms of length, tortuosity and radius for VesselVAE, minimum matching distance, coverage and \onenna{} for VesselGPT (scores from 3D point-cloud generation \citep{achlioptas2018,yang2019pointflow}), and point-cloud and graph distances for PartVessel. None of these scores is designed to separate fidelity from copying, nor are ours (Section~\ref{metrics}). Rankings under our scores need not transfer to point-cloud scores.

\section{Setup}\label{sec:setup}

A fair contest needs a reference that copies by design, the same patient splits for every method and a copying score beside fidelity; this section fixes all three.

\subsection{The resampler}\label{the-resampler}

\begin{figure}[!htb]\centering
\includegraphics[width=\linewidth]{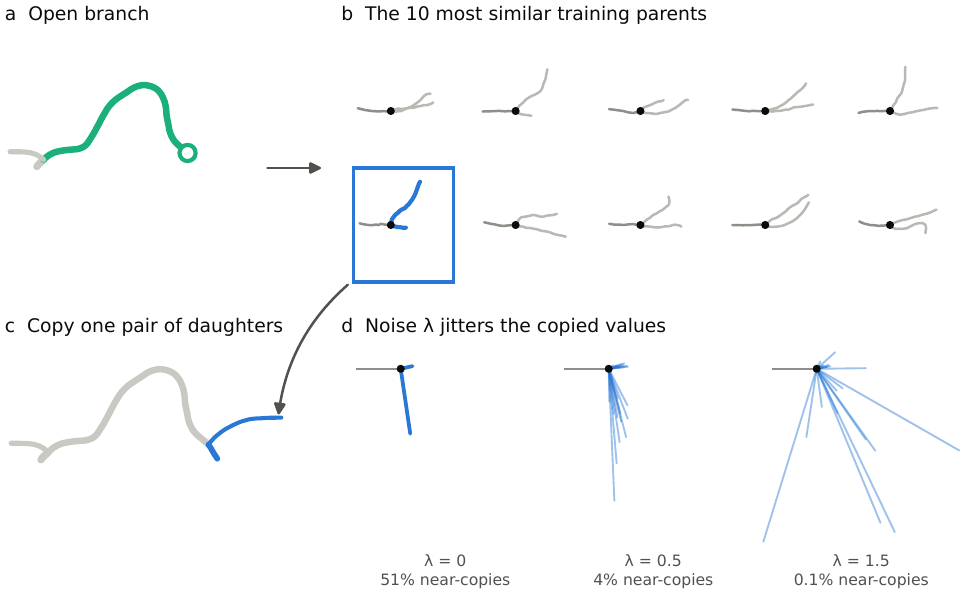}
\vspace{-7mm}
\caption{\textbf{The resampler grows a tree by copying the daughters of a similar training parent, and noise $\lambda$ sets how closely it copies.} (a) An open branch of the growing tree of Figure~\ref{fig:1}. (b) The $K = 10$ training bifurcations at the same depth whose parent is most similar in length, end radius and tortuosity; one is picked at random (boxed). (c) Its daughters are attached: lengths, tortuosities, angles, radius ratio and slope are copied. (d) Noise of size $\lambda$ (in training standard deviations) perturbs the copied values, shown for this daughter pair as straight segments (length and angle only), with the near-copy rate of branch records on the coronary test set. \label{fig:2}}
\vspace{-3mm}
\end{figure}

A nonparametric conditional resampler grows a tree by giving each open branch the daughters of a similar training parent (Figure~\ref{fig:2}). No parameters are optimised: the model is the training set and $K$. It suits the role of a reference for three reasons. It reproduces local branch statistics by construction, so a learned generator that does worse on them has not yet learned what a lookup gives for free. Its copying is explicit and can be turned down by a single noise level, so fidelity can be traced against copying. And it involves no training, so no question of budget or convergence arises for it.

\textbf{A one-hop branching process.} Each training bifurcation $j$ gives a key $k_j$ for the parent branch and a payload $\pi_j$ for its two daughters, with one split flag per daughter (whether it bifurcates again in its training tree). An open branch with key $k$ draws its daughters from
\[
\hat p(\pi \mid k) = \frac{1}{K} \sum_{j \in \mathcal N_K(k)} \delta_{\pi_j},
\]
where $\mathcal N_K(k)$ holds the $K$ training bifurcations at the same depth with keys closest to $k$ (over all depths, with depth as a fourth key coordinate, where a depth has fewer than $K$) and $\delta_{\pi_j}$ is a point mass. In words, the branch copies the payload $\pi_J$ of a similar training parent, $J$ uniform on $\mathcal N_K(k)$: a nonparametric one-hop (first-order Markov) branching process, or nearest-neighbour bootstrap \citep{lall1996} on trees.

\textbf{Noise.} As $\hat p$ only returns seen payloads, the resampler with noise level $\lambda$ perturbs every continuous value of $\pi_J$ in a transformed space $g$ (e.g.\ log length):
\[
g(\pi') = g(\pi_J) + \lambda\, \sigma_g\, \varepsilon,\qquad \varepsilon \sim \mathcal N(0, I),
\]
with $\sigma_g$ the training standard deviation of each transformed value. In words, copied daughters are nudged by about $\lambda$ such deviations (Figure~\ref{fig:2}d); split flags are never perturbed, though perturbed keys change which flags downstream lookups copy. The point masses become kernels (Gaussian in $g$, reflected or wrapped for angles), the smoothed bootstrap \citep{silverman1987}: $\lambda = 0$ copies, a large $\lambda$ keeps mainly the copied split flags, and sweeping $\lambda$ traces the fidelity-versus-copying curve (Section~\ref{sec:q3}).

\textbf{Generation.} A tree starts from a random training root, copied and perturbed like a payload (inlet radius included). An open branch draws daughters from $\hat p$ only if its copied split flag says so. Each daughter's key comes from its own copied, perturbed values at the next depth, so every draw sees only its parent. Trees above the branch cap are redrawn; the depth cap stops splitting (Appendix~\ref{app:spec-resampler}).

\textbf{Vessel features.} The key is $k = (d, \log L, \log r^{\mathrm{end}}, \tau)$: the parent's depth, log length, log end radius and tortuosity. Neighbours share the parent's depth and are ranked on the other three, standardised; this matching carries parent-daughter dependence. The payload is everything needed to draw both daughters: each daughter's length $L$, tortuosity $\tau$, radius slope $s$, start-radius ratio $\rho$ to the parent's end radius, bifurcation angle $\alpha$ and split flag $\beta \in \{0,1\}$, plus their azimuth difference $\Delta\phi$ (the angle between the two daughters around the parent's axis). A daughter's end radius follows from $\rho$ and $s$, clipped to the training range. Only these summaries are copied, never coordinates: 3D paths are drawn afresh with the copied length and tortuosity. $K = 10$ is untuned; coronary data get one resampler per artery (fallbacks and transforms: Appendix~\ref{app:repro}).

\subsection{Data and compared methods}\label{data-and-compared-methods}

\textbf{Coronary.} From the public CT angiography dataset ImageCAS \citep{zeng2023imagecas} we extracted in-house left and right centerline trees with radii for 546 patients (1,092 trees; Appendix~\ref{app:repro}), split by patient into 446 / 50 / 50 for every method; coronary results use the 100 test trees. Trees with expert annotations are pruned to the main named branches (left: left main, LAD, diagonals, circumflex, obtuse marginals; right: RCA, acute marginals, posterolateral and posterior descending branches), so they are small trees; the 114 trees without annotations are kept unpruned (Appendix~\ref{app:repro}).

\textbf{IntrA.} The public intracranial dataset IntrA \citep{yang2020intra}, as the centerline trees with radii distributed with VesselVAE \citep{vesselvae2023}, gives 785 trees from 72 patients, in 5 patient-grouped folds shared by all methods; ``$\pm$'' is the spread across folds, not a confidence interval.

\textbf{Controls and learned baselines.} The \emph{independent sampler} draws every branch quantity from its training distribution at that depth, ignoring parent-daughter dependence. Three controls are scored as if they were generators: the \emph{training copy} returns the training trees unchanged; the \emph{jittered copy} adds small Gaussian noise to their point positions (Appendix~\ref{app:repro}); and the \emph{real validation trees}, held out from scoring but used for model selection, are a real reference. The learned baselines, VesselGPT, VesselVAE and PartVessel, are retrained from the authors' code.

\subsection{Training budgets of the learned baselines}\label{training-budgets-of-the-learned-models}

Each learned baseline was trained with the authors' code and learning-rate schedule (stretched to the budget) under a ceiling of 72 hours per run, and every run ended before that ceiling, by early stopping, by divergence or, for PartVessel's second stage, at the authors' default of 4,000 epochs; checkpoints and decoding options were chosen on validation data, except PartVessel's second stage, which has no validation loss (Appendix~\ref{app:budgets}). Three facts bound the comparison (Appendix~\ref{app:earlier}): VesselVAE and VesselGPT stopped far short of the authors' default epoch counts; VesselVAE's configuration failed its IntrA anchor check (agreement with the authors' checkpoint); and PartVessel's preprocessing alone costs fidelity (its round-trip reference, real coronary test trees passed through and back, scores \onenna{} 0.855). Learned outputs are also rescaled and not artery-matched (Appendix~\ref{app:baselines}), although a resampler pooled across arteries scores like the per-artery one (Table~\ref{tab:E10}). Verdicts describe these retrains, not the methods (Appendix~\ref{app:baselines}).

\subsection{Metrics}\label{metrics}

Fidelity is closeness to \emph{held-out} trees; copying, to \emph{training} trees. One fixed pipeline turns every tree into 11 \emph{tree features}, whole-tree summaries such as branch count and total length, and into \emph{branch records}: 10 numbers per non-root branch describing it, its parent and its closest sibling. The branch record is the unit of copying because it is the unit the resampler draws: a daughter pair with the parent it was chosen for.

\textbf{\onenna{} (fidelity of whole trees).} Subsample the generated set $G$ to the size of the real test set $R$, pool them, and ask whether each tree's nearest tree-feature neighbour comes from its own set:
\[
\mathrm{1NNA}(G, R) = \frac{1}{|G| + |R|}\Big(\sum_{g \in G} \mathbb 1\big[\mathrm{NN}(g) \in G\big] + \sum_{r \in R} \mathbb 1\big[\mathrm{NN}(r) \in R\big]\Big).
\]
It is about 0.5 for indistinguishable sets and 1 for separated ones: a leave-one-out nearest-neighbour two-sample test \citep{lopezpaz2017,yang2019pointflow}. It judges trees at the scale of whole-tree summaries, while the next score judges them at the scale of single bifurcations.

\textbf{Dependence AUC (fidelity of parent-daughter relations).} A per-column empirical-CDF transform, fitted within each class (generated or real), normalises the unweighted branch-record marginals of both classes, so the score emphasises cross-feature dependence \citep{sklar1959}. A classifier separates generated from real branch records; we report its out-of-fold ROC AUC (grouped by real patient or generated tree; mean of 200 repeats, each averaging the five cross-validation folds on IntrA) on all 10 columns (the default) and on the six length, radius and tortuosity columns (Appendix~\ref{app:repro}). The score mainly reads relations, not values: on validation data two real samples score 0.53, while a positive control whose columns are permuted, destroying the relations but keeping every marginal, scores 0.91 (calibration and power: Appendix~\ref{app:dependence}). For both scores, closer to 0.5 is better. These two are \emph{our fidelity scores}; they are not point-cloud scores.

\begin{figure}[!htb]\centering
\includegraphics[width=\linewidth]{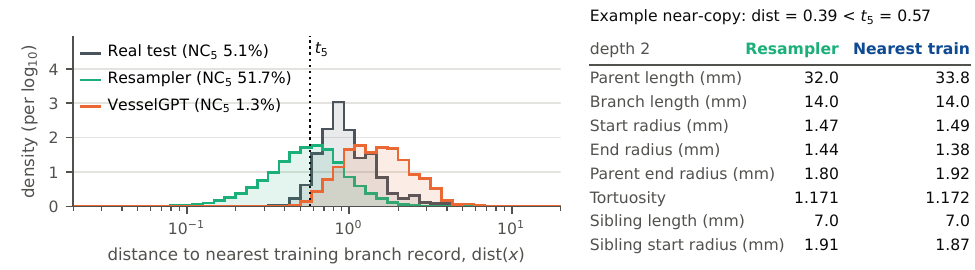}
\vspace{-4mm}
\caption{\textbf{About half of the resampler's branch records lie closer to a training record than 95\% of real test records do, against 1.3\% of VesselGPT's (coronary, $q = 5$).} Left: density of $\mathrm{dist}(x)$, the distance of a standardised branch record to its nearest training branch record, for real test, resampler and VesselGPT records; dotted line: $t_5$. Right: the resampler's median near-copy record ($\mathrm{dist} = 0.39 < t_5 = 0.57$, depth 2) beside its nearest training record, in physical units (taper not shown). \label{fig:ncexample}}
\end{figure}

\textbf{Near-copy rate (copying).} Let $\mathcal R_G$ be the branch records of $G$, $\mathrm{dist}(x)$ the distance from a standardised branch record $x$ to its nearest training branch record, and $t_q$ the $q$-th percentile of $\mathrm{dist}$ over real test branch records:
\[
\mathrm{NC}_q(G) = \frac{1}{|\mathcal R_G|}\sum_{x \in \mathcal R_G} \mathbb 1\big[\mathrm{dist}(x) < t_q\big]
\]
is the share of generated branch records closer to the training data than all but $q\%$ of real test branch records, which score about $q\%$ by construction (other held-out real trees only approximately; Section~\ref{sec:q3}); a rate far above $q\%$ flags excess proximity to training records, which warrants checking for reuse. We use $q = 5$ and check $q = 1$ and $10$ (Appendix~\ref{app:tables}). Figure~\ref{fig:ncexample} makes this concrete: in physical units, the resampler's median near-copy agrees with its nearest training record to within about 1.8\,mm in every length and 0.12\,mm in every radius, with the same tortuosity to two decimals, while only 1.3\% of VesselGPT's records fall below $t_5$, fewer than the 5\% of real test records.

Fidelity and copying can now be read side by side.

\section{The leaderboard}\label{sec:q1}

\begin{figure}[!htb]\centering
\includegraphics[width=\linewidth]{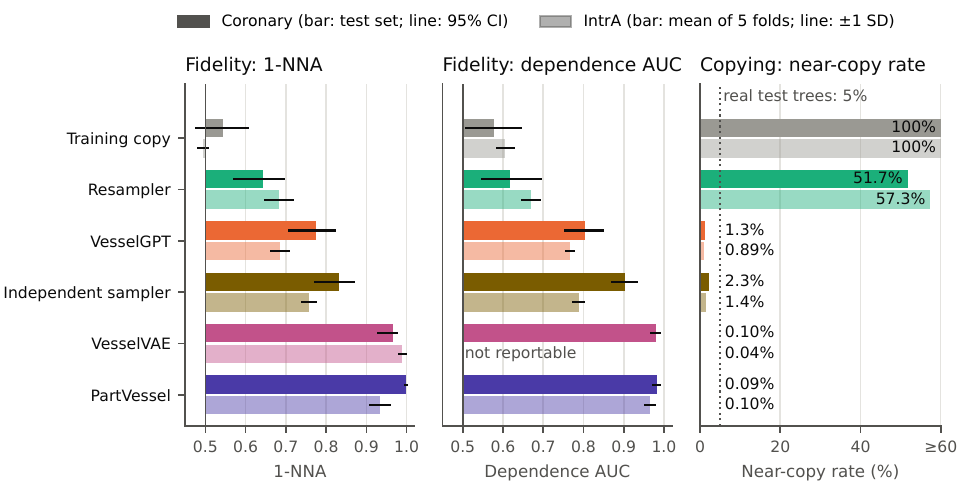}
\vspace{-9mm}
\caption{\textbf{The resampler leads the learned baselines on both fidelity scores while about half of its branch records are near-copies, and a training copy scores better still.} \onenna{} (left) and dependence AUC (middle) against real test trees (0.5 = indistinguishable; closer to 0.5 is better), and the near-copy rate at $q = 5$ (right; dotted: real test branch records, 5\% by construction). Dark bars: coronary test set, with 95\% CIs; light bars: IntrA, mean of 5 folds $\pm$ 1 SD. Rows are sorted by coronary \onenna{}; the training copy (a control) is on top. VesselVAE's IntrA dependence AUC is not reportable (Appendix~\ref{app:dependence}). The IntrA training copy's \onenna{} (0.494) lies below 0.5, so its bar is not visible. \label{fig:3}}
\end{figure}

We start where a reader of a generator paper starts, a fidelity leaderboard, and find a method with no parameter optimisation at the top of it (Figure~\ref{fig:3}; values in Table~\ref{tab:1}, Appendix~\ref{app:tables}).

Among the generators, the resampler has the best point estimate on both fidelity scores on both datasets. Only on IntrA \onenna{} is its lead over VesselGPT narrow (0.683 against 0.685). Even the independent sampler, which ignores the parent, is closer to real trees than VesselVAE and PartVessel on every reported score. That sampler needs nothing but the distribution of each branch quantity at each depth, so it is a natural floor; two of the three learned baselines, as retrained here, fall short of it.

This is surprising: a high rank is normally read as evidence that a generator has captured the distribution of whole trees. The resampler does not directly model dependencies beyond its local conditioning state: each draw sees only one parent's key. Yet \onenna{} judges 11 whole-tree summaries, and the resampler still matches them best among the generators.

A sceptical reader has two doubts. First, why would a nonparametric resampler win? Every daughter pair in a resampled tree carries the copied descriptors of a real daughter pair, chosen to suit its parent (Figure~\ref{fig:1}), so the tree inherits real branch lengths, angles and parent-daughter relations that a learned generator must produce itself. Additive tree features, such as branch count and total length, then come out realistic too. The random gallery shows typical output: the resampler's trees resemble the real ones, while VesselVAE's and PartVessel's visibly do not (Figure~\ref{fig:4}; IntrA trees in Appendix~\ref{app:galleries}). The independent sampler isolates what the parent adds: its coronary dependence AUC is 0.903 against the resampler's 0.618. Matching the parent is what makes the resampler's parent-daughter relations realistic.

\begin{figure}[!htb]\centering
\includegraphics[width=\linewidth]{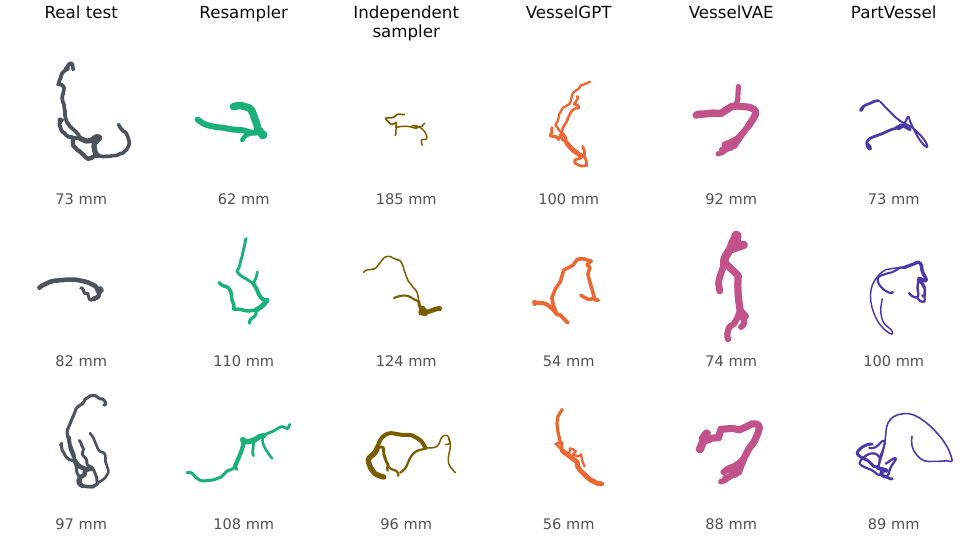}
\vspace{-4mm}
\caption{\textbf{Resampled coronary trees look like real ones, while VesselVAE's and PartVessel's visibly do not.} Three trees per column, drawn at random with a fixed seed (no selection) from the real test set and from each method, all with the same renderer; the number under each tree is its largest extent in mm. VesselVAE's trees are thick with few branches, and PartVessel's paths loop back on themselves. \label{fig:4}}
\vspace{-4mm}
\end{figure}

The second doubt is noise: a best point estimate is not a detected difference, and the coronary test set holds only 50 patients. We therefore test each pairing (Table~\ref{tab:2}, Appendix~\ref{app:tables}). On coronary data the resampler's lead is detected against every learned baseline on both scores; on IntrA, against PartVessel on both and VesselVAE on \onenna{}. Against VesselGPT on IntrA the comparison is inconclusive: the lead is detected in at most 1 of 5 folds, and equivalence is not shown. The test set also shaped the method, so these labels are descriptive (paired bootstrap with Holm adjustment \citep{holm1979}; decision rules: Appendix~\ref{app:dependence}).

\textbf{Part of the lead is built in.} The resampler conditions on parent length, end radius and tortuosity, and copies the branch quantities from which both scores' features are computed, so both scores are partly good for it by construction (Appendix~\ref{app:disclosures}). The lead also depends on which tree features \onenna{} reads: on scale-free tree features (branch count, depth, mean tortuosity), no \onenna{} difference from VesselGPT is detected, and VesselGPT's point estimate is the closer to 0.5; VesselVAE and PartVessel remain detectably worse (Table~\ref{tab:E10}, Appendix~\ref{app:disclosures}).

The resampler does lead these retrains of the learned baselines on our fidelity scores. Yet a verbatim training copy scores better still (Figure~\ref{fig:3}), so the next section asks how much of that lead is copying.

\section{What the leaderboard cannot see}\label{sec:q2}

The resampler tops the leaderboard, but it is built by copying; this section asks how much of its output is new and whether the fidelity scores can tell.

\textbf{Each tree is new; its branch records often are not.} About half of the resampler's branch records are near-copies of training branch records on both datasets (51.7\% coronary, 57.3\% IntrA), while the learned baselines stay at or below the 5\% of real held-out branch records (Figure~\ref{fig:3}, right). A lower rate is not better in itself: real validation trees reach 6.8\% (Figure~\ref{fig:concept}b). The ordering holds at $q = 1$ and $q = 10$ (Table~\ref{tab:E1}). Why half, and not all? A copied daughter keeps its own length, tortuosity, radius ratio and slope, and its sibling is copied with it, but its branch record also describes its parent, which in the generated tree only resembles the parent in the training tree. The closer the parent match, the more records are near-copies, which the sweep over $K$ below confirms.

Copying lives at the scale of a bifurcation, not of a tree. Applied to whole trees in the 11 tree features that \onenna{} uses (Appendix~\ref{app:repro}), the same near-copy rule flags only 6.3--7.2\% of the resampler's coronary trees per seed, barely above the 5\% of real test trees (Appendix~\ref{app:tables}). The reuse is visible only one level down.

\textbf{The fidelity scores do not notice.} Both scores ask whether generated trees look like new patients from the same population. A copy of a training patient is exactly that, so it passes (Section~\ref{sec:intro}). In our data, the training copy, which generates nothing, has the best \onenna{} and the best dependence AUC on both datasets (Figure~\ref{fig:3}). On their own, our fidelity scores cannot establish novelty.

Nor do the scores track the near-copy rate. Every $K$ copies daughters verbatim, but $K$ sets how closely the parent context matches: the coronary near-copy rate is 82\% at $K = 1$ and 40\% at $K = 30$, and the resampler with noise level $\lambda = 0.25$ reaches 14\%. Over this whole range both scores barely move, and their confidence intervals overlap (hollow markers and the right end of the curve in Figure~\ref{fig:6}). Removing most of the near-copies costs the resampler almost nothing on either score, so the scores do not appear to reward copying as such, but realistic branch values and parent-daughter relations, which copying supplies. One score summarises whole trees, the other parent-daughter relations; neither is built to see copying. Neither the training copy nor the resampler is trained, so this holds however the learned baselines were trained.

What drives the near-copy rate is the match between a copied daughter pair and its new parent. Conditioning on depth alone, which still copies every daughter pair verbatim but ignores the parent's length, radius and tortuosity, gives a coronary near-copy rate of 9.5\%, because the copied daughters rarely meet a parent like their source parent. It also loses fidelity, most clearly on the dependence AUC (0.768 against 0.613 for the curve's $\lambda = 0$ point, a second seed of the resampler; Table~\ref{tab:E3}): the parent context that makes records near-copies is the same context that makes parent-daughter relations realistic. Copying and fidelity are thus entangled in the resampler by design.

\textbf{What the data-copying test sees.} The data-copying test compares generated and held-out distances to the nearest training point; its score $C_T$ is negative when generated points lie closer to the training set (Appendix~\ref{app:repro}). The whole-tree version of the test of \citet{meehan2020} does not flag the resampler: its coronary score $C_T$ reads as underfitting, because a tree assembled from several patients' daughter pairs (Figure~\ref{fig:1}) lies farther from its nearest training tree than held-out trees do. The test is also underpowered on coronary trees: a half-copy control, exact training copies at the resampler's near-copy rate and the rest from the independent sampler, reaches only about $-1.1$, short of our copying cut-off of $-2$ (Table~\ref{tab:3}, Appendix~\ref{app:repro}). Applied to branch records, the test flags the resampler on both datasets, but it then uses the same records as the near-copy rate, so it confirms that rate rather than replicating it independently, and it is likely anti-conservative because branch records cluster within trees.

Each resampled tree is therefore new, but about half of its branch records are near-copies, and neither our fidelity scores nor a whole-tree copying test notices. The near-copy rate does, although it measures proximity of branch records, not every form of reuse (Section~\ref{sec:q3}). The fair question is then how the learned baselines compare with the resampler at the same near-copy rate.

\section{Comparing at the same near-copy rate}\label{sec:q3}

This section compares the learned baselines with the resampler at the same near-copy rate and probes the comparison with the controls.

\begin{figure}[!htb]\centering
\includegraphics[width=\linewidth]{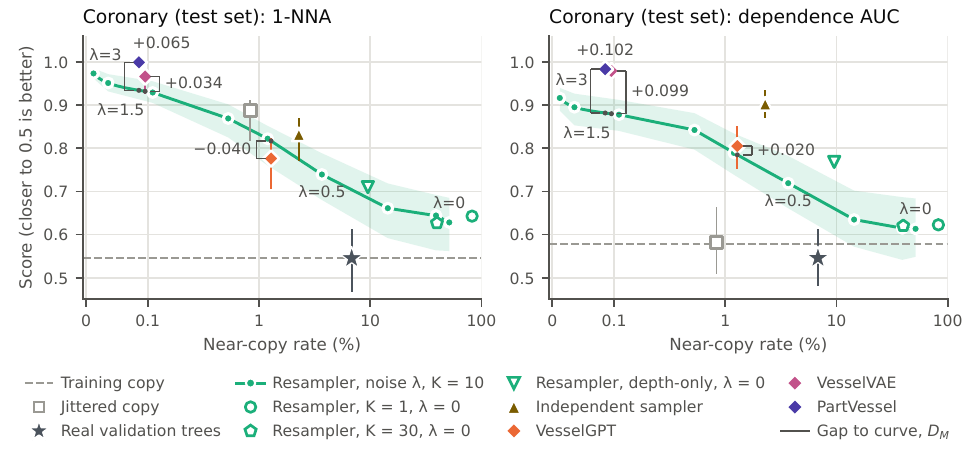}
\caption{\textbf{At its own near-copy rate no learned baseline lies below the resampler's curve, while real validation trees do on both scores and a jittered copy does on the dependence AUC (coronary test set).} \onenna{} (left) and dependence AUC (right) against the branch-record near-copy rate at $q = 5$ (log axis; zero rates at 0); closer to 0.5 is better. Line and shaded band: the resampler with noise level $\lambda$ ($K = 10$, $\lambda = 0$ to 3) with its 95\% band. Its $\lambda = 0$ point is a second seed of the resampler of Figure~\ref{fig:3}. Hollow markers at $\lambda = 0$: $K = 1$ (circle), $K = 30$ (pentagon) and depth-only conditioning (inverted triangle; Appendix~\ref{app:repro}). Hollow square: jittered copy; star: real validation trees; filled triangle: independent sampler; diamonds: the learned baselines, with 95\% bars; dashed line: training copy. Brackets: the gap to the curve $D_M$ at $q = 5$ of each learned baseline, labelled with its value (intervals in Table~\ref{tab:4}, Appendix~\ref{app:tables}); on the dependence AUC panel the VesselVAE marker is partly hidden behind PartVessel's. Intervals are conditional on the stored generated sets. IntrA: Figure~\ref{fig:curveintra}. \label{fig:6}}
\end{figure}

\textbf{Turning near-copies down.} Sweeping the noise level $\lambda$ of Section~\ref{the-resampler} from 0 to 3 traces the fidelity-versus-copying curve: the resampler's fidelity against its near-copy rate (Figure~\ref{fig:6}; IntrA: Figure~\ref{fig:curveintra}; grid: Appendix~\ref{app:disclosures}). An advantage over the curve is then not explained by branch-record proximity, though other reuse is not excluded.

\textbf{Gap to the curve.} A generator that copies as rarely as VesselGPT should be compared with a resampler that copies as rarely, not with one that copies half of its records. The curve supplies that reference: the score resampling reaches at a model's measured near-copy rate. For a model $M$ with near-copy rate $c_M$ and fidelity score $f_M$, let $\hat f(c_M)$ be the curve's score at the same near-copy rate, interpolated in $\log c$. The gap to the curve is
\[
D_M = f_M - \hat f(c_M).
\]
A negative $D_M$ means the model is closer to real trees than the resampler with noise at the same near-copy rate; a positive one means resampling with noise would have scored better. A 95\% interval wholly on one side of 0 places the model below or above the curve (interval construction: Appendix~\ref{app:repro}). The comparison matches only the thresholded branch-record proximity statistic, not the amount of reuse; the controls below test this.

\textbf{Reading Figure~\ref{fig:6}.} Start at the dashed line: the training copy copies every branch record and scores better than any generator. Moving left from the resampler at $\lambda = 0$, the curve is first flat: near-copies can be removed almost for free, whether by matching parents less closely ($K = 30$) or by a little noise ($\lambda = 0.25$ keeps 14\%). At the low rates where the learned baselines sit, it turns steep and fidelity worsens quickly (Tables~\ref{tab:E3} and~\ref{tab:E4}). Part of the steep rise may reflect leaving the data's support: radii are clipped and fewer trees are valid (0.68 at $\lambda = 3$ on coronary validation runs; Appendices~\ref{app:disclosures} and~\ref{app:spec-resampler}).

\textbf{The learned baselines.} No learned baseline lies below the curve on coronary data, and two lie above it; the brackets in Figure~\ref{fig:6} show each gap. VesselVAE and PartVessel are above the curve on the dependence AUC, where the resampler is strong partly by construction (Section~\ref{sec:q1}), and PartVessel is also above it on \onenna{} at $q = 5$. VesselGPT is the closest case. Of its coronary branch records, 1.3\% are near-copies, about as many as the resampler produces at noise level $\lambda = 0.75$ (1.2\%; Table~\ref{tab:E3}), and VesselGPT's two scores fall on either side of that resampler's scores. No difference is detected, but the interval on its \onenna{} gap still allows an advantage of about 0.12, so VesselGPT is inconclusive (Table~\ref{tab:4}, Appendix~\ref{app:tables}). On IntrA the picture is less clear: no learned baseline is below the curve in all folds, and VesselGPT is below it on \onenna{} in 2 of 5 (Table~\ref{tab:E8}). The comparison detects large defects but not small advantages. The independent sampler lies above the curve on the coronary dependence AUC, so a large defect is caught, but this is a positive control and not a power analysis: on coronary data the 95\% intervals on $D_M$ have half-widths of about 0.04 to 0.07 (Table~\ref{tab:4}, Appendix~\ref{app:tables}), so gaps of that size can go undetected.

\textbf{Perturbed copies lie below the curve.} A copy of the training trees with its points moved can lie below the curve, so a matched near-copy rate does not rule out copying. The jittered copy shows this. It keeps the topology and radii of every training tree and moves each point with Gaussian noise of 0.5\,mm. That is enough to lower its coronary near-copy rate to 0.83\%, fewer near-copies than VesselGPT has, yet it lies below the curve on the coronary dependence AUC and is never above it on IntrA (Table~\ref{tab:E9}). Stronger jitter of 1\,mm also lies below the curve on the dependence AUC, although it ruins the tree summaries (\onenna{} 0.998; at $q = 5$ only by extrapolation beyond the curve, Table~\ref{tab:E10}). So does a coronary copy displaced smoothly by 4\,mm (Table~\ref{tab:E12}). \onenna{} is not immune either. No coronary jitter level lies below the curve on it, but on IntrA a copy jittered by a quarter of the point spacing does so in all 5 folds (Table~\ref{tab:E11}). Matching the near-copy rate therefore certifies neither score against perturbed copies. The likely reason is that moving the points changes lengths and tortuosities enough to push branch records past the near-copy radius. The copied topology and radii, meanwhile, keep what the scores read: the parent-daughter relations for the dependence AUC and, under light jitter, the tree summaries for \onenna{}.

\textbf{Real trees lie below the curve too.} Real validation trees come from patients outside the training set, yet they lie below the curve on both coronary scores, so the curve does not bound attainable fidelity and a low near-copy rate is no goal in itself. Their \onenna{} is 0.545, against about 0.70 for the resampler at the same near-copy rate. On IntrA, their size-matched \onenna{} is below the curve in all 5 folds at $q = 5$ (Table~\ref{tab:E9}). Their near-copy rates are not exactly $q\%$, because the thresholds are set on the test records rather than on them, and trees from new patients still contain some branches that are nearly identical to training branches.

At the same near-copy rate, then, no learned baseline is detectably closer to real trees than the resampler, and lying below the curve informs without certifying: real trees and perturbed copies get there too.

\section{Discussion}\label{sec:discussion}\label{sec:limitations}

\textbf{What the evidence shows.} Fidelity scores computed against held-out trees cannot, by themselves, see reuse of training data. The resampler leads three retrained generators on our fidelity scores, detectably except against VesselGPT on IntrA or on scale-free tree features, while about half of its branch records are near-copies, and a training copy scores better still. Only the near-copy rate, not the whole-tree data-copying test, exposes this reuse. At the same near-copy rate, VesselVAE and PartVessel are worse on parent-daughter relations and VesselGPT is inconclusive. Copying is not bad in itself: a resampler is the right reference, and resampled trees may serve as data if users know they recombine training anatomy (only meshing was tested, Appendix~\ref{app:meshing}); the unmeasured risk is overstated local diversity, so copying must be visible.

\textbf{Limits.} Four kinds of limit qualify the comparison: the learned baselines may be undertrained, several choices favour one side or the other, the resampler is an easy reference to beat, and the near-copy rate does not catch all reuse. On training, VesselVAE and VesselGPT stopped short of their authors' default epoch counts, VesselVAE's configuration failed its anchor check, and PartVessel's preprocessing alone costs fidelity (Section~\ref{training-budgets-of-the-learned-models}). Some choices favour the resampler. Both scores give it a built-in advantage (Section~\ref{sec:q1}), and on coronary data it is also fit per artery and scored on more trees, while the learned sets are not artery-matched. Other choices favour the learned baselines. Their checkpoints and decoding options were chosen on validation data, while the resampler has no tuned parameter, and on IntrA the resampler outputs binary trees (Appendix~\ref{app:fairness}). The resampler is easy to beat because it uses only one hop of parent context; because its noise leaves the copied split flags unchanged, so topology changes only through downstream lookups and cap rejection; and because at high noise it leaves the data's support and yields fewer valid trees (Appendix~\ref{app:disclosures}). Finally, as the perturbed copies show (Section~\ref{sec:q3}), the near-copy rate misses some reuse, which also limits the comparison at matched near-copy rates.

The evidence itself is limited. The study is exploratory: the test sets were used throughout, also to design the resampler and the evaluation, so every verdict is descriptive; IntrA folds share training data, and there is no multiplicity correction (Appendix~\ref{app:disclosures}). VesselGPT remains inconclusive: on IntrA, on scale-free features and against the curve, where its coronary gap intervals are about 0.15 wide; the dependence AUC also has low power on IntrA (noise threshold 0.178 against coronary 0.094; Appendix~\ref{app:dependence}). The verdicts cover two datasets of small trees and these retrains, not the methods; code and coronary trees are unreleased (Appendix~\ref{app:spec}). The near-copy rate is stable when the point representation is re-sampled; a denser curve is not tested (Appendix~\ref{app:disclosures}). Three findings could change the comparative answers: a learned baseline, trained to its published schedule, that lies below the curve by more than jittered copies do and approaches the real reference; a retrain of VesselVAE that passes its anchor check; or a PartVessel preprocessing without a fidelity cost.

\textbf{How to use the diagnostic.} What would beating resampling look like? A generator that had learned how real trees are built would approach held-out real trees near their own near-copy rate, not merely fall below the curve, which real trees and a jittered copy also do. The near-copy rate is therefore a diagnostic to report with its controls, not a test to pass. The argument does not depend on vessels: any fidelity score computed only against held-out data passes a training copy. A comparison on one split and one scoring pipeline should score a verbatim training copy and the resampler, since a table won by either cannot establish novelty, and report the branch-record near-copy rate at $q = 1$, 5 and 10 next to every fidelity score (Section~\ref{metrics}). It should then place the generator against the fidelity-versus-copying curve at its own rate, together with jittered copies at several levels and held-out real trees, with an interval on $D_M$ conditional on the fixed generated sets (Section~\ref{sec:q3}, Appendix~\ref{app:repro}). Finally, it should state each generator's training budget, anchor check and preprocessing cost (Appendix~\ref{app:baselines}); otherwise a gap describes a retrain, not a method.

Fidelity scores will rightly remain the main way to judge generated anatomy. But a fidelity score answers only one of the two questions a user of synthetic trees asks. The other, how much of the output is the training data again, needs its own number, measured where reuse happens. With its controls, the near-copy rate makes that second question visible.

\FloatBarrier
\clearpage
\bibliography{refs}
\bibliographystyle{tmlr}

\clearpage
\appendix
\section{Scope of the evidence}\label{app:disclosures}

\textbf{Status.} The study is exploratory and descriptive. The coronary and IntrA test sets were used throughout, including in the design of the resampler and of the evaluation, so there is no untouched hold-out set and no confirmatory test. There is no multiplicity correction across analyses (Holm adjustment is used only inside the head-to-head rules of Appendix~\ref{app:dependence}), and every label, verdict and interval describes these data. Fixed before the analysis they govern was scored on test data: the near-copy thresholds, the head-to-head decision rules (Appendix~\ref{app:dependence}), the curve grid up to $\lambda = 1.5$ and the rule that extends it, the reading rule for the control analyses, and the scale-free feature subset. Set with test results available (of the reported methods, or of other configurations of the learned baselines and of an unpublished learned model): the resampler, the dependence test and the near-copy rate, which were fixed before the reported baseline configurations were trained; the equivalence margin; the gap to the curve $D_M$ and its intervals; the data-copying test $C_T$; the application of the grid extension ($\lambda = 2.0$ and 3.0); and the control analyses with their perturbation levels (Tables~\ref{tab:E9}--\ref{tab:E12}). The real validation trees were used for model selection and in the grid-extension rule: they are held out from scoring, not untouched.

\textbf{Advantages built into the comparison.} The resampler conditions on parent length, end radius and tortuosity, and copies the branch quantities from which both scores' features are computed, so its \onenna{} and dependence AUC are partly good by construction. Its other advantages and disadvantages are listed in Appendix~\ref{app:baselines}. On IntrA it outputs binary trees (at nodes with three or more children it is fit on a random pair); PartVessel was trained on trees with extra children removed but is scored for copying against the full fold training set.

\textbf{Margin, grid, $D_M$ and $C_T$.} The TOST margin is the coronary training copy's \onenna{} minus 0.5; it was set with the per-fold IntrA \onenna{} estimates known and is also used for $D_M$, so TOST results are descriptive. The curve grid up to $\lambda = 1.5$ was fixed before scoring (its points 0.75 and 1.5 chosen on validation data only), and each level was scored once on test data. The grid is extended by a fixed rule that reads only the resampler's validation near-copy rates and the learned baselines' reported rates; it was applied with the test results of the grid up to $\lambda = 1.5$ available. Under it, 2.0 and 3.0 are always added, then 4.0 and 5.0 until the highest level's coronary validation rate is at or below the learned baselines' lowest coronary test rate at every threshold (0, 0.086\% and 0.086\% at $q = 1, 5, 10$). It stops at $\lambda = 3.0$, whose validation rate is 0 at all three thresholds; IntrA uses the same levels. Coronary verdicts for VesselVAE and PartVessel at $q = 5$ and 10 rely on the extended grid. The paired comparison of $D_M$ uses the extended grid to bracket VesselVAE and PartVessel, treats floor rates separately and, on IntrA, resamples the curve and the near-copy rate (conditional on the fixed generated sets and seeds); a reading by point estimates alone would change 17 of 36 comparable verdicts. $C_T$ was set with the near-copy results known but before its values on real data were computed.

\textbf{Control analyses.} The reading rule is: if a perturbed training copy lies below the curve on a score, matching the measured branch-record near-copy rate does not certify that score against perturbed copies, and the Q3 conclusions on that score are restricted. On coronary data the jittered copy (SD 0.5\,mm; near-copy rate 0.83\% at $q = 5$) is below the curve on the dependence AUC at every threshold ($D_M = -0.23$, 95\% CI $[-0.32, -0.15]$, at $q = 5$), with no difference detected on \onenna{}. On IntrA it is below the curve on the dependence AUC (all features) in 2--3 of 5 folds and above in none; most of its \onenna{} comparisons are not determined because its near-copy rate sits at the resolution floor. The further coronary levels (SD 0.1, 0.25 and 1\,mm) were chosen with the 0.5\,mm result known and use different seeds from it. The 1\,mm copy is below the curve on the dependence AUC at $q = 1$ and, by extrapolation beyond the last curve point, at $q = 5$ and 10; no coronary level is below on \onenna{}; the two mildest are not determined, as their near-copy rates exceed the curve's range. On IntrA the further levels are SD 0.02, 0.05 and 0.2 coordinate units (point spacing 0.2), with common random numbers across levels (Table~\ref{tab:E11}). The 0.05 copy keeps 25\% near-copies at $q = 5$ and an \onenna{} of 0.526, and lies below the curve on \onenna{} in all 5 folds at $q = 1$ and 5 ($D_M$ from $-0.20$ to $-0.15$, every interval below 0), at rates inside each fold's curve. On the dependence AUC (all features) only the 0.2 copy is below in all 5 folds, at $q = 10$ and at the near-copy floor (0\% near-copies, compared with the curve points that also have none). The IntrA \onenna{} result rests on this one level, a quarter of the point spacing. Smoothly displaced copies (Gaussian displacement fields with a correlation length of 5 point spacings, amplitudes from 0.1 to 4 spacings; the levels above 1 spacing are set by a rule that reads validation data only) stay near-copies far longer than jittered ones at the same amplitude (coronary, $q = 5$: 98.8\% at 0.5\,mm and 89.5\% at 1\,mm; Table~\ref{tab:E12}). On coronary data the copy displaced by 4\,mm (near-copy rate 0.43, 0.83 and 1.58\% at $q = 1, 5, 10$) lies below the curve on the dependence AUC at every threshold (0.592 against 0.76--0.81; $D_M$ from $-0.22$ to $-0.17$) and on its feature subset, while lying above it on \onenna{} (0.976); this rests on that single level. The 2\,mm copy (18.9\%) shows no difference on either score. On IntrA no smooth level lies below the curve in all 5 folds: at 0.4 units \onenna{} is below in 4 of 5 folds at $q = 1$ and 5, and at 0.8 units the dependence AUC is below in 4 of 5 folds at $q = 5$ and 10, the fifth fold not determined. Matching the measured near-copy rate therefore certifies neither score against perturbed copies: the dependence AUC on both datasets and under both kinds of perturbation, \onenna{} on IntrA. No tested level, jittered or smooth, lies below the curve on the scale-free \onenna{}.

\textbf{Point re-sampling.} Re-sampling the training copy's polylines at 0.5 to 2 times the point spacing leaves it a near-copy (at least 99.63\% of branch records on coronary data and 98.70\% on IntrA, at every threshold and fold). Re-sampling the real test trees at 1.5 and 2 times the spacing moves the near-copy thresholds by a factor of 0.98--1.04, keeps 4.8--6.1\% of the re-sampled test records below the original $q = 5$ threshold, and changes the $q = 5$ rate of no curve point or learned baseline with at least 10 near-copied records by more than 16\% (coronary $\lambda = 0$: 51.4 to 50.0\%; VesselGPT: 1.28 to 1.16\%). The thresholds and rates are therefore stable under this change of representation; the verdicts were not recomputed at the re-sampled thresholds, and the stability cut-offs were set with these re-sampling results available.

\textbf{Real validation trees.} Scored against test data as if they were a generator (Figure~\ref{fig:6}), their measured near-copy rate does not sit at $q$: coronary 2.0, 6.8 and 12.9\% at $q = 1, 5, 10$, IntrA 3.5--11.4\% across folds at $q = 5$. On coronary data they are below the curve on both scores at every threshold, by about 0.16 on \onenna{} and 0.13 on the dependence AUC at $q = 5$. On IntrA, \onenna{} is computed against a test subsample of the validation set's size: 0.46--0.545 per fold against 0.64--0.69 for the curve, below it in 5 of 5 folds at $q = 5$. Validation trees are more patient-clustered than that subsample, which biases their \onenna{} upward, so this verdict is conservative. Their IntrA dependence AUC on all features is below the curve in 2 of 5 folds and not otherwise distinguished from it; on the feature subset no fold is below.

\textbf{Validity and cap rejection.} On coronary validation runs the share of valid trees is 0.98, 0.92, 0.87 and 0.68 at $\lambda = 0, 1.5, 2, 3$, and cap rejection stays below 1\% of attempts at every $\lambda$. The validity flag does not include branch collisions, whose share rises from 0.11 to 0.42 over the same range. On IntrA the flag does not measure generator validity, since real IntrA trees also score 0.17--0.28 (one-point branches are part of those data).

\textbf{Pooled resampler and scale-free \onenna{}.} At $\lambda = 0$, a coronary resampler pooled across arteries without artery labels differs from the per-artery one by $+$0.008 [$-$0.057, $+$0.073] on \onenna{} and $+$0.004 [$-$0.078, $+$0.084] on the dependence AUC. The scale-free subset is branch count, depth and mean tortuosity. On coronary data VesselGPT's distance from 0.5 minus that of the resampler at the curve's $\lambda = 0$ point is $-$0.005 [$-$0.067, $+$0.095] on it (11 features: $+$0.148 [$+$0.065, $+$0.225]; Table~\ref{tab:2} uses the main resampler sets and the opposite sign); VesselVAE and PartVessel remain detectably worse. No learned baseline's coronary verdict against the curve changes class between the two feature sets; the independent sampler (above to no difference, at $q = 5$ and 10), the 0.5\,mm jittered copy (no difference to above) and the real validation trees (below to no difference) do. On IntrA, against the curve's point at $\lambda = 0$ (11-feature fold mean 0.693, from the curve's generated sets, not those of Table~\ref{tab:1}), the scale-free \onenna{} fold means are 0.561 for the resampler, 0.543 for VesselGPT, 0.934 for VesselVAE and 0.814 for PartVessel: VesselVAE and PartVessel are worse than the resampler in all 5 folds (paired 95\% intervals above 0), while VesselGPT is not distinguished from it in any fold, as on all 11 features against that point. Against the scale-free curve, VesselGPT's two folds below the curve become no difference at $q = 10$ (at $q = 5$, one no difference and one not determined), and VesselVAE's and PartVessel's not-determined folds become above.

\textbf{Baselines.} Every run ended before the 72-hour ceiling, by early stopping, by divergence or, for PartVessel's second stage, at the authors' default of 4,000 epochs. VesselGPT and VesselVAE stopped far short of the authors' default epoch counts; PartVessel's first stage stopped at 12,000--30,000 epochs, around the 20,000 its paper reports. The VesselVAE anchor failed. Other configurations of all three baselines were also trained and scored on the same test sets (Appendix~\ref{app:earlier}), and several steps depart from the authors' code (Appendix~\ref{app:departures}). No claim is made that these are faithful reproductions of the published models.

\textbf{Not covered.} The meshing check uses n = 20 trees per method, below its planned sample size (Appendix~\ref{app:meshing}), and the IntrA noise threshold of the dependence test was measured on one fold and applied to all. The study does not test a densified curve with rare-event bounds, or two-level and per-seed sensitivities. On IntrA it runs no pooled resampler (IntrA has no artery labels).

\section{Baselines}\label{app:baselines}

\subsection{Budgets and stopping}\label{budgets-and-stopping}\label{app:budgets}

\textbf{VesselGPT.} Default in the authors' code: 500,000 stage-1 and 50,000 stage-2 epochs. The paper reports about 3 h + 24 h of training and no epoch counts. Our ceiling: 24 h + 48 h per training seed. Both stages stopped early, at 9.8--16.5 h and 13.3--14.6 h; at most about 660 stage-1 and 2,140 stage-2 epochs would have fit in the ceiling.

\textbf{VesselVAE.} Default in the authors' code: 200,000 epochs. The paper reports about 12 h of training and no epoch counts. Our ceiling: 72 h on one CPU core, which is faster than a GPU for this recursive model. Runs ended by early stopping or divergence at 26--29 h; the best coronary checkpoints are at epochs 850--1,220, and about 8,400 epochs would have fit.

\textbf{PartVessel.} Default in the authors' code: 40,000 stage-1 and 4,000 stage-2 epochs. The paper reports 20,000 + 2,000 epochs. Our ceiling: stage 1 until 40,000 epochs or 60 h, then stage 2, within 72 h. Stage 1 stopped early at 14,000--22,000 epochs on coronary data and 12,000--30,000 on IntrA, around the published schedule.

Published schedules are from the papers \citep{vesselgpt, vesselvae2023, partvessel}. Learning-rate schedules are those of the authors' code, stretched to the budget. Coronary: 3 training seeds per baseline, 200 trees per seed. IntrA: one training seed per fold (VesselVAE best epochs 185--270), 300 trees per set. Decoding choices on validation \onenna{}: VesselGPT five options (top-k 50 at temperatures 0.3, 0.7, 1.0; 4-beam sampling at 0.3, 1.0), VesselVAE two node caps.

\begin{figure}[!htb]\centering
\includegraphics[width=\linewidth]{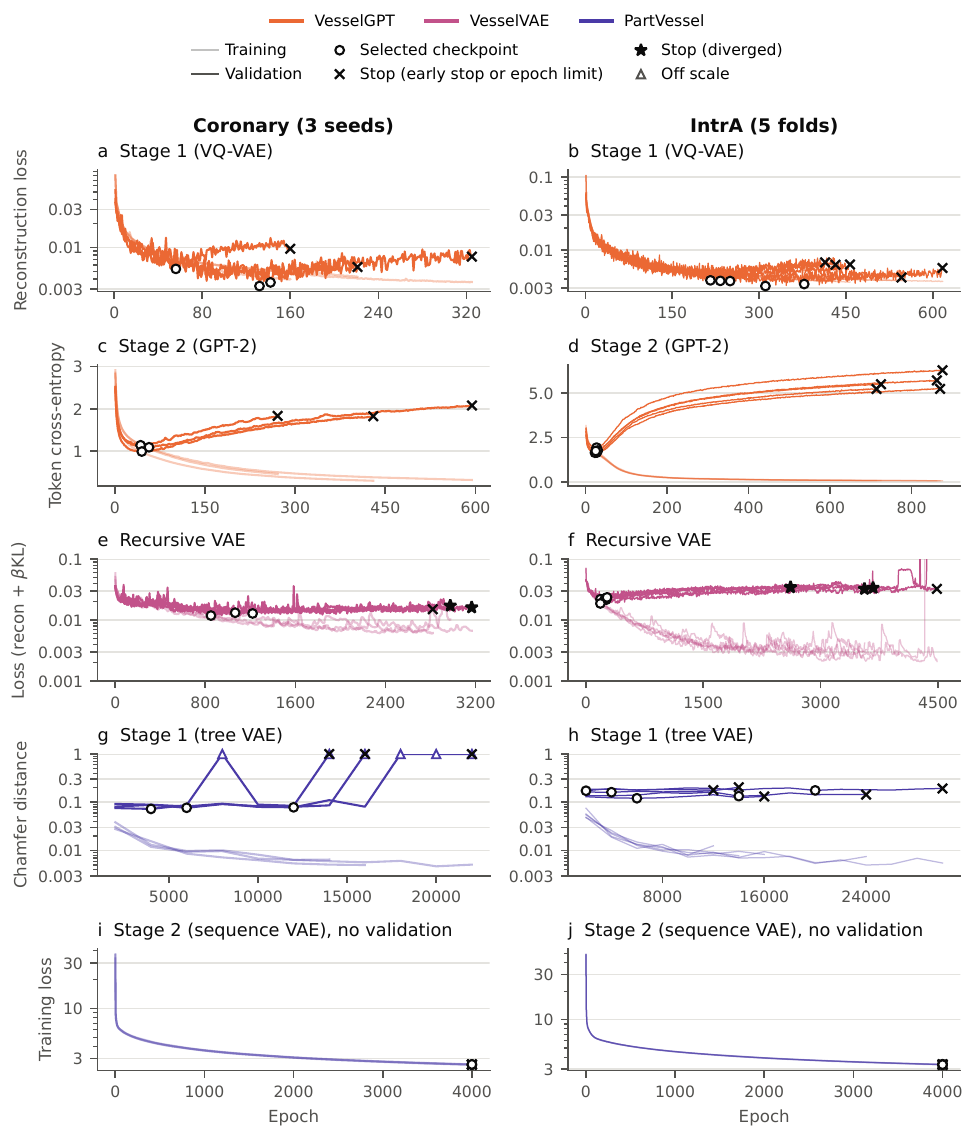}
\caption{\textbf{Validation losses stop improving early while training losses keep falling, so the selected checkpoints are not limited by the training budget on these criteria.} Left: coronary (three training seeds); right: IntrA (one run per fold). Rows: VesselGPT stage 1 (VQ-VAE reconstruction loss) and stage 2 (token cross-entropy); VesselVAE (reconstruction plus weighted KL term); PartVessel stage 1 (tree-VAE Chamfer distance, evaluated every 2,000 epochs) and stage 2 (training loss only, no validation). Dark curves: each run's own selection criterion (PartVessel stage 1: Chamfer distance on the authors' internal held-out split). VesselVAE training loss is a running median over 25 epochs (batches of 2 spike), clipped above 0.1.\label{fig:training}}
\end{figure}

Figure~\ref{fig:training} shows the training and validation curves of every reported run, with the selected checkpoint and the reason each run stopped. VesselGPT (both stages) and PartVessel stage 1 stopped by early stopping in every run. VesselGPT's stage-2 validation cross-entropy is lowest at epochs 23--57 and then rises while the training loss falls. VesselVAE's validation minimum comes early (IntrA epochs 185--270), and every divergence happened long after the selected checkpoint.

\subsection{Departures from the authors' code}\label{departures-from-the-authors-code}\label{app:departures}

Outputs are rescaled to the extent of a random training tree with the same branch count, because the models do not store per-tree normalisation. The same 889 coronary training trees are used for all three (three trees below 20 mm extent, which PartVessel's preprocessing drops, are dropped for all). VesselVAE and VesselGPT select checkpoints on validation loss (the authors' code partly on training loss). VesselVAE: two seeds use checkpoints from before divergence, and sampling is re-seeded after the authors' code fixes its own seed at import. PartVessel: the log-variance is clamped to prevent NaN failures.

\subsection{Other configurations, anchors and round-trip references}\label{earlier-configurations-anchors-and-ceilings}\label{app:earlier}

\textbf{Other configurations.} Other configurations of each learned baseline were also trained on the same data and scored on the same coronary test set. None enters any comparison, and each scores worse than the reported one: \onenna{} 0.984 for VesselGPT (its dependence AUC not reliable), 0.997 for VesselVAE (dependence AUC 0.996) and 1.000 for PartVessel (0.992), against 0.776, 0.966 and 0.999 for the reported configurations.

\textbf{Round trips.} A round trip passes real test trees through a baseline's preprocessing and back, with no model involved, so it measures the distortion of the conversion alone. Each round-trip tree is a perturbed copy of a tree in the reference set, so a conversion that loses little scores at or below 0.5. Through PartVessel's preprocessing, real test trees score \onenna{} 0.855 on coronary data and 0.666 $\pm$ 0.008 on IntrA: the conversion alone costs PartVessel fidelity.

\textbf{VesselVAE anchor.} The authors' checkpoint (latent size 32, selected on training loss) has a stored loss that matches training on normalised IntrA segments (extent 0.25 to 1.0); its training patients were not published and may include our test patients, so its 0.984 is not held out. On segments in millimetres its loss is non-finite. Its outputs carry no physical scale and are sized by our common rescaling rule. An anchor retrain of our configuration (three seeds on the same data, separate from the per-fold models of Table~\ref{tab:1}, which score 0.989) should land within a tolerance of this checkpoint; it scores 0.998 and misses the tolerance (gap 0.0139 against 0.0123), so the anchor check fails. The score therefore describes the checkpoint as run in this evaluation, not the model as evaluated by its authors. The representation alone does not explain the score: real IntrA test trees passed through VesselVAE's preprocessing and back score 0.428 $\pm$ 0.030. The authors' demo decoding accepts a leaf only once two bifurcations lie on its root path, and all 3,000 of its trees have 7 to 11 branches (real median 3). At matched extent their radii are about twice the real ones. Our retrain (seed 0 checked) shows the same structure.

\subsection{Fairness in both directions}\label{fairness-in-both-directions}\label{app:fairness}

Some choices favour the resampler: it is a test-aware reference (Appendix~\ref{app:disclosures}); on coronary data it is fit per artery and scored on about 1,000 trees per seed against 200, while the learned sets are not artery-matched. Others favour the learned baselines: their checkpoints and decoding options were chosen on validation data, while the resampler has no tuned parameter ($K = 10$ is untuned; a validation sweep over $K$ is descriptive only), and on IntrA the resampler outputs binary trees. Every method is scored by the same code on the same test sets, with the same training split defining the near-copy distances.
\FloatBarrier

\section{The dependence test and the decision rules}\label{app:dependence}

\textbf{Calibration.} On validation data: two real samples 0.53 {[}0.45, 0.61{]}; a column-permuted positive control 0.91; shuffle sanity 0.49--0.50. Synthetic Gaussian copula: 0.51 with no change, 0.60 when a correlation of 0.8 is removed, 0.49 for a monotone change of a marginal. Permutation null for two real coronary samples (200 permutations of whole patients between the training and test roles, 40 repeats each; Appendix~\ref{app:spec-inference}): mean 0.548 {[}0.523, 0.578{]} against an observed 0.574 at the same 40 repeats (0.577 with the 200 repeats of Table~\ref{tab:1}), one-sided p = 0.04. Two samples of exchangeable patients thus score about 0.55, so most of the real-versus-real excess over 0.5 is pipeline optimism; the true split sits at the upper edge of the null, either a small genuine difference between training and test patients (about 0.03) or chance. Weighting variants leave conclusions unchanged. On validation data the resampler gives 0.58--0.59 for $K$ from 1 to 30 and depth-only conditioning 0.76: the test responds to the strength of one-hop conditioning.

\textbf{Reporting conditions.} Each repeat resamples test patients, subsamples the stored generated trees and refits the empirical-CDF transform and the classifier within grouped cross-validation (real trees by patient, generated trees by tree); the full procedure is in Appendix~\ref{app:spec-inference}. A dependence AUC is reported only if a calibration gate passes (the 2.5 percentile of the column-permuted training copy above the 97.5 percentile of the training copy, on both feature sets) and the method's sanity AUC on shuffled branch records is at most 0.55. The gate passes on coronary test data and in 5/5 IntrA folds; shuffle sanity is 0.492--0.497 (coronary) and 0.497--0.503 (IntrA) for reported methods; VesselVAE fails it on IntrA. Noise thresholds (twice the largest SD of paired differences between generators that should be indistinguishable, on validation data; Appendix~\ref{app:spec-inference}): coronary 0.094 (all columns) / 0.098 (subset), IntrA 0.178 / 0.156.

\textbf{Decision rules} (their status: Appendix~\ref{app:disclosures}). \emph{\onenna{}:} the paired bootstrap 95\% CI of $\Delta\lvert \onenna{} - 0.5 \rvert$ (resampler minus baseline, $B = 1000$) lies below 0 and the Holm-adjusted p is below 0.05. \emph{Dependence:} the gate passes; the resampler is lower on both feature sets; each CI excludes 0; each $\lvert\Delta\mathrm{AUC}\rvert$ reaches the noise threshold; shuffle sanity at most 0.55; Holm-adjusted p below 0.05. Holm is applied within a dataset across the three baselines, per metric (per fold on IntrA); on IntrA a baseline counts as beaten only in 5/5 folds. \emph{Equivalence} (IntrA, VesselGPT, \onenna{}; margin set with the per-fold IntrA \onenna{} estimates available, Appendix~\ref{app:disclosures}): TOST at $\pm$0.045, requiring both the pooled-bootstrap and the across-fold 90\% intervals inside the margin. The smallest margins the data support are 0.034 (pooled) and 0.048 (across folds).
\section{Reproducibility details}\label{app:repro}

This appendix summarises settings; feature definitions, the resampler's algorithm and constants, the inference procedures and the IntrA splits are specified in Appendix~\ref{app:spec}.

\textbf{Resampler.} If fewer than $K$ training bifurcations exist at a depth, the search runs over all depths with depth as a fourth standardised key coordinate. Branch paths are smooth random curves with the copied length and tortuosity, and radius profiles are linear and clipped to the training range. Trees above the branch cap are rejected and redrawn; a branch at the depth cap does not split (caps, path model and calibration table: Appendix~\ref{app:spec-resampler}). Each draw looks only at the open branch's own key (its depth, length, end radius and tortuosity), never at its ancestors or at the rest of the tree, so the resampler models one hop of context and nothing longer. Depth-only conditioning (``depth-only'' in Figure~\ref{fig:6}) draws uniformly among all bifurcations at the same depth.

\textbf{Noise transforms.} With independent $\varepsilon \sim \mathcal N(0,1)$ per value: $L' = L\, e^{\lambda \sigma_{\log L} \varepsilon}$; $\tau' = 1 + (\tau - 1)\, e^{\lambda \sigma_{\log(\tau-1)} \varepsilon}$; $s' = s + \lambda \sigma_s \varepsilon$; $\rho' = \rho\, e^{\lambda \sigma_{\log \rho} \varepsilon}$; $\alpha'$ is $\alpha + \lambda \sigma_\alpha \varepsilon$ reflected into $[0^\circ, 180^\circ]$; $\Delta\phi' = (\Delta\phi + \lambda \sigma_{\Delta\phi} \varepsilon) \bmod 2\pi$; the root's $L$, $\tau$ and $s$ use the same transforms with SDs over training roots, and its inlet radius is perturbed multiplicatively; $\beta$ is copied unchanged. SDs are computed per model on the training set. Curve sets: coronary 3 seeds $\times$ 1,000 trees per point; IntrA 300 trees per fold per point.

\textbf{Near-copy rate.} Branch records (Table~\ref{tab:H2}) are standardised with training means and SDs, missing values set to the training mean; nearest training branch record in Euclidean distance; thresholds at the 1st, 5th and 10th percentiles of test-to-training distances (coronary 0.452, 0.574, 0.633; IntrA per fold). Coronary branch record counts: training 4,864, test 554. The tree-level variant (Section~\ref{sec:q2}) applies the same rule to the 11 standardised tree features, with the 5th percentile of the real test trees' nearest-training distances.

\textbf{\onenna{}.} 11 tree features (Table~\ref{tab:H1}) z-scored with training statistics; each generated set subsampled without replacement to the reference size; intervals from a bootstrap over reference patients ($B = 1000$) that also draws the stored seed sets with replacement; copies of a resampled tree are never each other's neighbours (Appendix~\ref{app:spec-inference}).

\textbf{Dependence test.} Within-class empirical-CDF transform; gradient-boosted classifier (200 iterations); 5-fold cross-validation grouped by patient (real) or tree (generated; training and jittered copies by source patient); 200 repeats; no early stopping (Appendix~\ref{app:spec-inference}). Subset columns: lengths of branch, parent and sibling, start radii of branch and sibling, and tortuosity.

\textbf{Jittered copy.} Training trees with Gaussian noise on point positions (SD 0.5 mm coronary, 0.1 IntrA coordinate units), radii unchanged; coronary 3 sets of 200 trees (100 per artery), IntrA 300 trees per fold. Further coronary levels (SD 0.1, 0.25, 1\,mm): Table~\ref{tab:E10}.

\textbf{Paired curve comparison (Section~\ref{sec:q3}).} The 95\% interval of the gap to the curve $D_M$ resamples test patients, the near-copy threshold and trees within the stored sets (paired patient bootstrap \citep{efron1979} for \onenna{}, with seed sets resampled among themselves; the 200 stored repeats for the dependence AUC). An interval wholly on one side of 0 places a model below or above the curve; otherwise no difference is detected. For \onenna{} on all 11 features a negative $D_M$ reads as closer to real trees when both scores are at least 0.5; the only lower ones, real validation folds of 0.46 or more, remain closer to 0.5 than the curve (scale-free \onenna{}: Table~\ref{tab:E10}). Near-copy rates are floored at $0.5/N_{\min}$ for plotting and interpolation only; zero observed near-copies do not establish a zero population rate; $N_{\min}$ is the smallest branch record count among the curve points and compared methods. Interpolation is linear in $\log_{10}$ rate between the first $\lambda$-consecutive pair of curve points that bracket the model's rate; without a bracketing pair the nearest point is used and the comparison is flagged ``outside the curve's copy range'' (then only ``below the curve (extrapolated)'' or ``not determined''; a model that copies more than every curve point is always ``not determined''). If the model's floored rate equals the floor together with two or more curve points, the curve value is the range of those points, $D_M$ the midpoint of the two differences, and the 95\% interval runs from the 2.5\% point of the difference to the range's maximum to the 97.5\% point of the difference to its minimum; the comparison is flagged ``tie'' and is ``not determined'' unless that interval excludes 0. A flag applies if its status holds at the point estimate or in more than 5\% of replicates. \onenna{}: 1,000 patient-bootstrap replicates shared by the model and all curve points, each recomputing the near-copy threshold from the resampled test branch records. Dependence AUC: the 200 stored repeats, whose real resamples are identical across methods. Equivalence is assessed for \onenna{} only, by TOST at $\pm$0.045 (descriptive only; Appendix~\ref{app:disclosures}) (coronary: unflagged and the 90\% interval inside the margin; IntrA: every fold's 90\% interval and the across-fold interval inside). No multiplicity correction. The interval does not include interpolation error between curve points; the $\lambda = 2.0$ and $3.0$ sets use seeds independent of the other curve points; for a model with no near-copies, only the resampled threshold moves its near-copy rate. Intervals are conditional on the stored generated sets and seeds (step-by-step procedure: Appendix~\ref{app:spec-inference}).

\textbf{Data-copying test (Section~\ref{sec:q2}).} $C_T$ \citep{meehan2020}: each generated and held-out test point gets its distance to the nearest point of the full training set; the space is split into k-means cells fitted on the training set, with $k = \lfloor n_{\text{test}}/20 \rfloor$ (coronary trees 5, branch records 27; IntrA trees 7, branch records 21--23); in each cell with at least 20 generated points a Mann-Whitney statistic $Z_U$ \citep{mannwhitney1947} compares the generated and test distances, and $C_T$ is the average of the cell values weighted by each cell's share of test points. Generated sets are subsampled to a fixed size (coronary 500 trees, 2,000 branch records; IntrA 250 trees, 600 branch records), 20 repeats with their own k-means seed and subsample. Controls: training copy; half-copy (exact training draws at the resampler's $q = 5$ rate, the rest from the independent sampler); real validation trees (not subsampled). Our reading (not part of the original test): $C_T \le -2$ copying, $\ge +2$ underfitting; ``detects copying'' also requires $C_T$ below the independent sampler's in every repeat (coronary) or fold (IntrA); ``underpowered'' if the half-copy control is not read as copying. Sensitivity runs at about $k/2$ and $2k$ and with the authors' reference code never make the tree level read copying for the resampler, and its branch-record $C_T$ stays at or below $-3.74$.

\textbf{Coronary pruning.} Coronary trees were pruned with per-patient expert annotations to the main named branches (left: left main, LAD, diagonals, circumflex, obtuse marginals; right: RCA, acute marginals, posterolateral and posterior descending branches). Septal, ramus and sinoatrial-nodal branches and unlabelled terminal branches were removed, branches left with a single child were merged into their parent, and the 114 trees without annotations were kept unpruned.

\textbf{IntrA trees and folds.} The 785 trees are the VMTK centerline trees of IntrA segments distributed with the VesselVAE code, at 0.2 point spacing (coordinate units). 5 patient-grouped folds over 72 patients: training 543--561 trees, validation 68--83 trees (7 patients), test 156--159 trees (14--15 patients) per fold.

\textbf{Validity under noise.} With noise, more target lengths fall below half a point spacing and give one-point branches, so on validation runs the share of valid generated trees (definition: Appendix~\ref{app:spec-resampler}) falls from 0.98 at $\lambda = 0$ to 0.92 at $\lambda = 1.5$ (larger $\lambda$: Appendix~\ref{app:spec-resampler}). Separately, lengths beyond the calibration table use its nearest row and radii are clipped to the training range.

\section{Full tables}\label{app:tables}

\begin{table}[tbp]\centering
\caption{\textbf{Among the generators, the resampler has the best fidelity point estimates on both datasets while about half of its branch records are near-copies; the training copy scores better still.} Point estimates; coronary: 100 test trees, IntrA: mean over 5 folds. Near-copy rate at $q = 5$. The training copy is a control, not a generator. VesselVAE's IntrA dependence AUC is not reportable: it fails the sanity check on shuffled branch records (Appendix~\ref{app:dependence}). Intervals and fold spreads are in Appendix Tables~\ref{tab:E0a} and \ref{tab:E0b}. \label{tab:1}}
\begin{tabular*}{\linewidth}{@{\extracolsep{\fill}}lcccccc@{}}
\toprule
\multirow{2}{*}{Method} & \multicolumn{3}{c}{Coronary} & \multicolumn{3}{c}{IntrA} \\
\cmidrule(lr){2-4} \cmidrule(lr){5-7}
 & \onenna{} & Dep. AUC & Near-copy & \onenna{} & Dep. AUC & Near-copy \\
\midrule
\emph{Training copy} & \emph{0.545} & \emph{0.577} & \emph{100\%} & \emph{0.494} & \emph{0.606} & \emph{100\%} \\
\textbf{Resampler} & \textbf{0.643} & \textbf{0.618} & 51.7\% & \textbf{0.683} & \textbf{0.669} & 57.3\% \\
Independent sampler & 0.832 & 0.903 & 2.3\% & 0.758 & 0.788 & 1.4\% \\
VesselGPT & 0.776 & 0.805 & 1.3\% & 0.685 & 0.767 & 0.9\% \\
VesselVAE & 0.966 & 0.979 & 0.10\% & 0.989 & not reportable & 0.04\% \\
PartVessel & 0.999 & 0.983 & 0.09\% & 0.933 & 0.965 & 0.10\% \\
\bottomrule
\end{tabular*}
\end{table}

\begin{table}[tbp]\centering
\caption{\textbf{The resampler's lead is detected against every learned baseline on coronary data, and on IntrA against PartVessel and VesselVAE (\onenna{}) but not VesselGPT.} Negative favours the resampler. For \onenna{} the difference is in $\lvert \onenna{} - 0.5 \rvert$; coronary $\Delta$ is a bootstrap mean of paired differences, not a difference of Table~\ref{tab:1} values. Coronary: paired bootstrap 95\% CI and Holm-adjusted p (dependence AUC: from the dependence test's 200 repeats). Dependence differences must also reach a noise threshold (all columns / subset: coronary 0.094 / 0.098, IntrA 0.178 / 0.156; Appendix~\ref{app:dependence}). IntrA: mean $\pm$ SD over folds, and folds (of 5) meeting the criterion. \label{tab:2}}
\begin{tabular*}{\linewidth}{@{\extracolsep{\fill}}llccccc@{}}
\toprule
\multirow{2}{*}{Baseline} & \multirow{2}{*}{Score} & \multicolumn{3}{c}{Coronary} & \multicolumn{2}{c}{IntrA} \\
\cmidrule(lr){3-5} \cmidrule(lr){6-7}
 &  & $\Delta$ & 95\% CI & Holm p & $\Delta$, mean $\pm$ SD & Folds \\
\midrule
 \multirow{2}{*}{VesselGPT} & \onenna{} & $-$0.136 & {[}$-$0.213, $-$0.063{]} & 0.003 & $-$0.007 $\pm$ 0.043 & 1/5  \\

& Dep. AUC & $-$0.187 & {[}$-$0.264, $-$0.100{]} & 0.030 & $-$0.097 $\pm$ 0.031 & 0/5  \\
\addlinespace
 \multirow{2}{*}{VesselVAE} & \onenna{} & $-$0.326 & {[}$-$0.387, $-$0.260{]} & 0.003 & $-$0.311 $\pm$ 0.034 & 5/5  \\

& Dep. AUC & $-$0.361 & {[}$-$0.432, $-$0.285{]} & 0.015 & not reportable &  \\
\addlinespace
 \multirow{2}{*}{PartVessel} & \onenna{} & $-$0.370 & {[}$-$0.432, $-$0.307{]} & 0.003 & $-$0.243 $\pm$ 0.022 & 5/5  \\

& Dep. AUC & $-$0.365 & {[}$-$0.440, $-$0.290{]} & 0.015 & $-$0.296 $\pm$ 0.022 & 5/5  \\
\bottomrule
\end{tabular*}
\end{table}

\begin{table}[tbp]\centering
\caption{\textbf{The whole-tree data-copying test does not flag the resampler; only its branch-record version does.} Coronary: mean (SD) over 20 repeats. IntrA: mean $\pm$ SD of the 5 fold means. The half-copy control holds 52\% (coronary) and 57\% (IntrA) exact training copies. Full results: Table~\ref{tab:E6}. \label{tab:3}}
\begin{tabular*}{\linewidth}{@{\extracolsep{\fill}}lcccc@{}}
\toprule
\multirow{2}{*}{Method} & \multicolumn{2}{c}{Whole trees} & \multicolumn{2}{c}{Branch records} \\
\cmidrule(lr){2-3} \cmidrule(lr){4-5}
 & Coronary & IntrA & Coronary & IntrA \\
\midrule
Training copy & $-$7.94 (0.08) & $-$7.38 $\pm$ 0.33 & $-$7.67 (0.07) & $-$7.87 $\pm$ 0.32  \\

Half-copy control & $-$1.10 (0.14) & $-$2.81 $\pm$ 0.15 & $-$2.44 (0.15) & $-$2.75 $\pm$ 0.06  \\

Resampler & +2.61 (0.20) & +1.39 $\pm$ 0.60 & $-$4.96 (0.09) & $-$5.30 $\pm$ 0.45  \\

Learned baselines (range) & +5.26 to +6.79 & +2.86 to +7.49 & +3.36 to +6.10 & +3.97 to +7.84  \\
\bottomrule
\end{tabular*}
\end{table}

\begin{table}[tbp]\centering
\caption{\textbf{At $q = 5$ no learned baseline lies below the curve, and VesselVAE and PartVessel lie above it on the coronary dependence AUC.} $D_M$ as defined in the text; dependence AUC uses all columns. Coronary: point estimate with 95\% CI; near-copy rates are in Table~\ref{tab:1}. IntrA: verdict across the 5 folds by the rule of Table~\ref{tab:E8}, which gives the fold counts; ``no diff.'' may include folds below or above the curve (VesselGPT \onenna{}: 2 of 5 below; independent sampler dep. AUC: 4 of 5 above) (n.d.: not determined; n/r: not reportable). \label{tab:4}}
\begin{tabular*}{\linewidth}{@{\extracolsep{\fill}}lccccll@{}}
\toprule
\multirow{2}{*}{Method} & \multicolumn{4}{c}{Coronary} & \multicolumn{2}{c}{IntrA} \\
\cmidrule(lr){2-5} \cmidrule(lr){6-7}
 & \onenna{} & 95\% CI & Dep. AUC & 95\% CI & \onenna{} & Dep. AUC \\
\midrule
Independent sampler & +0.058 & {[}+0.002, +0.103{]} & +0.154 & {[}+0.101, +0.203{]} & no diff. & no diff.  \\

VesselGPT & $-$0.040 & {[}$-$0.119, +0.026{]} & +0.020 & {[}$-$0.058, +0.087{]} & no diff. & mixed  \\

VesselVAE & +0.034 & {[}$-$0.033, +0.064{]} & +0.099 & {[}+0.050, +0.132{]} & mixed & n/r  \\

PartVessel & +0.065 & {[}+0.018, +0.113{]} & +0.102 & {[}+0.056, +0.139{]} & n.d. & above  \\
\bottomrule
\end{tabular*}
\end{table}

\begin{figure}[tbp]\centering
\includegraphics[width=\linewidth]{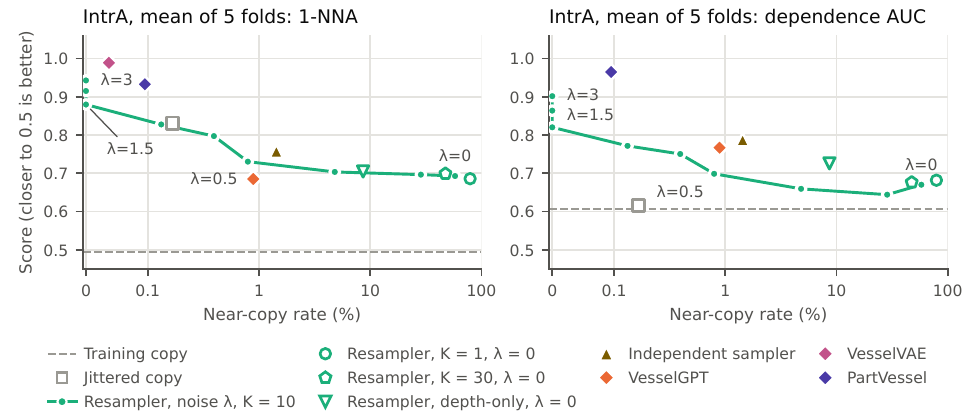}
\caption{\textbf{On IntrA VesselGPT's mean \onenna{} lies below the curve, yet no learned baseline is below it in all five folds.} As Figure~\ref{fig:6}, for IntrA: \onenna{} (left) and dependence AUC (right) against the branch-record near-copy rate at $q = 5$ (log axis; zero rates at 0). Each fold has its own curve; points are fold means, without intervals (per-fold verdicts: Table~\ref{tab:E8}). Only the IntrA training copy's \onenna{} (0.494) lies below 0.5. The real validation trees are not plotted: their size-matched \onenna{} is compared with a curve recomputed at the validation set's size, not with the plotted one (Table~\ref{tab:E9}). \label{fig:curveintra}}
\end{figure}

\begin{table}[tbp]\centering
\caption{\textbf{On coronary data the resampler's 95\% intervals do not overlap any learned baseline's, but overlap the training copy's.} Table~\ref{tab:1} with uncertainty, coronary. 100 test trees; 95\% bootstrap CIs. \label{tab:E0a}}
\begin{tabular*}{\linewidth}{@{\extracolsep{\fill}}lccccc@{}}
\toprule
\multirow{2}{*}{Method} & \multicolumn{2}{c}{\onenna{}} & \multicolumn{2}{c}{Dep. AUC} & \multirow{2}{*}{Near-copy} \\
\cmidrule(lr){2-3} \cmidrule(lr){4-5}
 & Estimate & 95\% CI & Estimate & 95\% CI &  \\
\midrule
\emph{Training copy} & \emph{0.545} & {[}0.477, 0.607{]} & \emph{0.577} & {[}0.507, 0.644{]} & \emph{100\%} \\
\textbf{Resampler} & \textbf{0.643} & {[}0.570, 0.695{]} & \textbf{0.618} & {[}0.548, 0.695{]} & 51.7\% \\
Independent sampler & 0.832 & {[}0.772, 0.868{]} & 0.903 & {[}0.872, 0.933{]} & 2.3\% \\
VesselGPT & 0.776 & {[}0.708, 0.822{]} & 0.805 & {[}0.754, 0.849{]} & 1.3\% \\
VesselVAE & 0.966 & {[}0.928, 0.977{]} & 0.979 & {[}0.968, 0.989{]} & 0.10\% \\
PartVessel & 0.999 & {[}0.997, 1.000{]} & 0.983 & {[}0.973, 0.991{]} & 0.09\% \\
\bottomrule
\end{tabular*}
\end{table}

\begin{table}[tbp]\centering
\caption{\textbf{On IntrA the resampler and VesselGPT have nearly equal \onenna{} (0.683 and 0.685), while VesselVAE and PartVessel trail far behind.} Table~\ref{tab:1} with uncertainty, IntrA. Mean $\pm$ SD across 5 folds (spreads, not CIs). \label{tab:E0b}}
\begin{tabular*}{\linewidth}{@{\extracolsep{\fill}}lccc@{}}
\toprule
Method & \onenna{} & Dep. AUC & Near-copy \\
\midrule
\emph{Training copy} & \emph{0.494} $\pm$ 0.012 & \emph{0.606} $\pm$ 0.021 & \emph{100\%} \\
\textbf{Resampler} & \textbf{0.683} $\pm$ 0.035 & \textbf{0.669} $\pm$ 0.022 & 57.3 $\pm$ 2.4\% \\
Independent sampler & 0.758 $\pm$ 0.018 & 0.788 $\pm$ 0.013 & 1.4 $\pm$ 0.6\% \\
VesselGPT & 0.685 $\pm$ 0.022 & 0.767 $\pm$ 0.010 & 0.9 $\pm$ 1.0\% \\
VesselVAE & 0.989 $\pm$ 0.009 & not reportable & 0.04 $\pm$ 0.02\% \\
PartVessel & 0.933 $\pm$ 0.025 & 0.965 $\pm$ 0.013 & 0.10 $\pm$ 0.18\% \\
\bottomrule
\end{tabular*}
\end{table}

\begin{table}[tbp]\centering
\caption{\textbf{At every threshold the resampler's near-copy rate is far above the real test records' rate, and every learned baseline's is below that rate.} Near-copy rate (\%) at three thresholds. Real test branch records $\approx$ 1 / 5 / 10\% by construction. IntrA: mean $\pm$ SD across folds. \label{tab:E1}}
\begin{tabular*}{\linewidth}{@{\extracolsep{\fill}}lcccccc@{}}
\toprule
\multirow{2}{*}{Method} & \multicolumn{3}{c}{Coronary} & \multicolumn{3}{c}{IntrA} \\
\cmidrule(lr){2-4} \cmidrule(lr){5-7}
 & q1 & q5 & q10 & q1 & q5 & q10 \\
\midrule
Jittered copy & 0.28 & 0.83 & 1.5 & 0.07 $\pm$ 0.11 & 0.17 $\pm$ 0.16 & 0.59 $\pm$ 0.33 \\
Resampler & 34.4 & 51.7 & 59.3 & 50.3 $\pm$ 1.9 & 57.3 $\pm$ 2.4 & 61.3 $\pm$ 2.7 \\
Independent sampler & 0.45 & 2.3 & 3.9 & 0.26 $\pm$ 0.17 & 1.4 $\pm$ 0.6 & 2.9 $\pm$ 1.4 \\
VesselGPT & 0.16 & 1.3 & 2.6 & 0.11 $\pm$ 0.08 & 0.9 $\pm$ 1.0 & 1.8 $\pm$ 1.7 \\
VesselVAE & 0.04 & 0.10 & 0.17 & 0.00 $\pm$ 0.00 & 0.04 $\pm$ 0.02 & 0.10 $\pm$ 0.04 \\
PartVessel & 0.00 & 0.09 & 0.09 & 0.00 $\pm$ 0.00 & 0.10 $\pm$ 0.18 & 0.21 $\pm$ 0.23 \\
\bottomrule
\end{tabular*}
\end{table}

Per fold, VesselGPT's IntrA rate at $q = 5$ is 2.7, 0.6, 0.2, 0.5 and 0.6\%. Tree-level near-copy rate (coronary, per seed): resampler 6.3, 7.2 and 7.1\%; independent sampler 0.5, 0.5 and 0.9\%; VesselGPT 1.5, 3.5 and 2.5\%.

\textbf{Subset dependence AUC.} Coronary {[}95\% CI{]}: training copy 0.543 {[}0.467, 0.621{]}, resampler 0.567 {[}0.498, 0.647{]}, independent sampler 0.800 {[}0.752, 0.846{]}, VesselGPT 0.672 {[}0.603, 0.740{]}, VesselVAE 0.880 {[}0.843, 0.911{]}, PartVessel 0.854 {[}0.808, 0.891{]}. Resampler minus baseline: VesselGPT $-$0.106 {[}$-$0.198, $-$0.016{]}, VesselVAE $-$0.314 {[}$-$0.383, $-$0.238{]}, PartVessel $-$0.288 {[}$-$0.363, $-$0.210{]}. IntrA: training copy 0.566 $\pm$ 0.015, resampler 0.623 $\pm$ 0.017, independent sampler 0.724 $\pm$ 0.024, VesselGPT 0.665 $\pm$ 0.019, PartVessel 0.835 $\pm$ 0.029.

\begin{table}[tbp]\centering
\caption{\textbf{On coronary data, noise first removes most near-copies at little cost in fidelity, then fidelity worsens quickly at the low rates where the learned baselines sit.} Coronary curve (95\% CIs). \label{tab:E3}}
\begin{tabular*}{\linewidth}{@{\extracolsep{\fill}}lcccccc@{}}
\toprule
\multirow{2}{*}{Setting} & \multirow{2}{*}{\onenna{}} & \multicolumn{2}{c}{Dep. AUC} & \multicolumn{3}{c}{Near-copy} \\
\cmidrule(lr){3-4} \cmidrule(lr){5-7}
 &  & all & subset & q1 & q5 & q10 \\
\midrule
\multicolumn{7}{@{}l}{\emph{Noise $\lambda$, K = 10}} \\

$\lambda$ = 0 & 0.628 {[}0.562, 0.685{]} & 0.613 {[}0.548, 0.684{]} & 0.567 {[}0.491, 0.633{]} & 34.2\% & 51.4\% & 59.1\%  \\

$\lambda$ = 0.1 & 0.644 {[}0.563, 0.692{]} & 0.614 {[}0.541, 0.687{]} & 0.563 {[}0.492, 0.636{]} & 21.3\% & 39.1\% & 47.8\%  \\

$\lambda$ = 0.25 & 0.661 {[}0.592, 0.718{]} & 0.635 {[}0.572, 0.702{]} & 0.558 {[}0.498, 0.630{]} & 5.3\% & 14.4\% & 20.7\%  \\

$\lambda$ = 0.5 & 0.739 {[}0.680, 0.788{]} & 0.719 {[}0.661, 0.782{]} & 0.609 {[}0.542, 0.687{]} & 0.9\% & 3.7\% & 6.2\%  \\

$\lambda$ = 0.75 & 0.822 {[}0.760, 0.852{]} & 0.789 {[}0.738, 0.834{]} & 0.666 {[}0.604, 0.724{]} & 0.3\% & 1.2\% & 2.1\%  \\

$\lambda$ = 1.0 & 0.869 {[}0.825, 0.902{]} & 0.842 {[}0.797, 0.881{]} & 0.731 {[}0.674, 0.788{]} & 0.1\% & 0.5\% & 1.0\%  \\

$\lambda$ = 1.5 & 0.929 {[}0.905, 0.955{]} & 0.878 {[}0.840, 0.912{]} & 0.765 {[}0.706, 0.818{]} & 0.0\% & 0.1\% & 0.2\%  \\

$\lambda$ = 2.0 & 0.951 {[}0.932, 0.972{]} & 0.895 {[}0.853, 0.926{]} & 0.780 {[}0.727, 0.825{]} & 0.00\% & 0.04\% & 0.11\%  \\

$\lambda$ = 3.0 & 0.973 {[}0.962, 0.985{]} & 0.916 {[}0.887, 0.940{]} & 0.795 {[}0.744, 0.836{]} & 0.00\% & 0.01\% & 0.03\%  \\
\addlinespace
\multicolumn{7}{@{}l}{\emph{Neighbours, $\lambda$ = 0}} \\

$K = 1$ & 0.643 {[}0.560, 0.692{]} & 0.622 {[}0.555, 0.694{]} & 0.575 {[}0.496, 0.645{]} & 69.5\% & 82.4\% & 86.9\%  \\

$K = 30$ & 0.627 {[}0.557, 0.688{]} & 0.620 {[}0.556, 0.692{]} & 0.572 {[}0.503, 0.634{]} & 23.1\% & 39.8\% & 48.0\%  \\

depth-only & 0.710 {[}0.635, 0.762{]} & 0.768 {[}0.700, 0.830{]} & 0.670 {[}0.605, 0.737{]} & 4.1\% & 9.5\% & 13.4\%  \\
\bottomrule
\end{tabular*}
\end{table}

A fresh-seed $\lambda = 0$ run against the main resampler sets (51.4\% vs 51.7\% copying, \onenna{} 0.628 vs 0.643, AUC 0.613 vs 0.618) shows the size of seed variability. The path, including $\lambda = 2.0$ and $3.0$ (set by the rule in Appendix~\ref{app:disclosures}, with independent seeds, and shown with two decimals because their rates are below 0.1\%), is monotone in near-copy rate at all three thresholds.

\begin{table}[tbp]\centering
\caption{\textbf{IntrA shows the same shape: fidelity changes little while most near-copies are removed, then worsens quickly at low rates.} IntrA curve (mean $\pm$ SD across folds). \label{tab:E4}}
\begin{tabular*}{\linewidth}{@{\extracolsep{\fill}}lcccccc@{}}
\toprule
\multirow{2}{*}{Setting} & \multirow{2}{*}{\onenna{}} & \multicolumn{2}{c}{Dep. AUC} & \multicolumn{3}{c}{Near-copy} \\
\cmidrule(lr){3-4} \cmidrule(lr){5-7}
 &  & all & subset & q1 & q5 & q10 \\
\midrule
\multicolumn{7}{@{}l}{\emph{Noise $\lambda$, K = 10}} \\

$\lambda$ = 0 & 0.693 $\pm$ 0.035 & 0.670 $\pm$ 0.021 & 0.625 $\pm$ 0.008 & 49.9 $\pm$ 4.6\% & 58.0 $\pm$ 4.1\% & 62.0 $\pm$ 4.4\%  \\

$\lambda$ = 0.1 & 0.696 $\pm$ 0.031 & 0.644 $\pm$ 0.018 & 0.597 $\pm$ 0.017 & 14.1 $\pm$ 4.4\% & 28.6 $\pm$ 4.7\% & 39.2 $\pm$ 5.7\%  \\

$\lambda$ = 0.25 & 0.703 $\pm$ 0.030 & 0.659 $\pm$ 0.027 & 0.602 $\pm$ 0.013 & 1.2 $\pm$ 0.5\% & 4.8 $\pm$ 2.3\% & 8.9 $\pm$ 3.5\%  \\

$\lambda$ = 0.5 & 0.730 $\pm$ 0.019 & 0.699 $\pm$ 0.027 & 0.616 $\pm$ 0.031 & 0.1 $\pm$ 0.2\% & 0.8 $\pm$ 0.7\% & 1.8 $\pm$ 1.3\%  \\

$\lambda$ = 0.75 & 0.797 $\pm$ 0.023 & 0.750 $\pm$ 0.034 & 0.664 $\pm$ 0.046 & 0.1 $\pm$ 0.2\% & 0.4 $\pm$ 0.5\% & 0.8 $\pm$ 0.8\%  \\

$\lambda$ = 1.0 & 0.828 $\pm$ 0.024 & 0.772 $\pm$ 0.023 & 0.689 $\pm$ 0.028 & 0.0 $\pm$ 0.1\% & 0.1 $\pm$ 0.1\% & 0.2 $\pm$ 0.2\%  \\

$\lambda$ = 1.5 & 0.880 $\pm$ 0.015 & 0.820 $\pm$ 0.020 & 0.720 $\pm$ 0.032 & 0.0 $\pm$ 0.0\% & 0.0 $\pm$ 0.0\% & 0.1 $\pm$ 0.1\%  \\

$\lambda$ = 2.0 & 0.915 $\pm$ 0.019 & 0.864 $\pm$ 0.018 & 0.747 $\pm$ 0.030 & 0.0 $\pm$ 0.0\% & 0.0 $\pm$ 0.0\% & 0.0 $\pm$ 0.0\%  \\

$\lambda$ = 3.0 & 0.943 $\pm$ 0.015 & 0.902 $\pm$ 0.013 & 0.773 $\pm$ 0.018 & 0.0 $\pm$ 0.0\% & 0.0 $\pm$ 0.0\% & 0.0 $\pm$ 0.0\%  \\
\addlinespace
\multicolumn{7}{@{}l}{\emph{Neighbours, $\lambda$ = 0}} \\

$K = 1$ & 0.686 $\pm$ 0.046 & 0.682 $\pm$ 0.021 & 0.633 $\pm$ 0.010 & 72.6 $\pm$ 4.2\% & 79.1 $\pm$ 3.3\% & 82.1 $\pm$ 2.6\%  \\

$K = 30$ & 0.699 $\pm$ 0.025 & 0.676 $\pm$ 0.015 & 0.630 $\pm$ 0.011 & 39.1 $\pm$ 3.9\% & 47.5 $\pm$ 3.0\% & 52.1 $\pm$ 3.0\%  \\

depth-only & 0.705 $\pm$ 0.028 & 0.726 $\pm$ 0.013 & 0.664 $\pm$ 0.021 & 4.4 $\pm$ 1.6\% & 8.6 $\pm$ 2.8\% & 12.9 $\pm$ 4.6\%  \\
\bottomrule
\end{tabular*}
\end{table}

On IntrA no fold has a near-copy at $\lambda = 2.0$ or $3.0$. The path is not monotone in folds 2--4 at $\lambda \ge 0.5$, where 0--2 near-copy branch records out of about 750 are counted.

\begin{table}[tbp]\centering
\caption{\textbf{On IntrA \onenna{}, the resampler's lead is detected against VesselVAE and PartVessel in every fold, and against VesselGPT in one fold of five.} IntrA head-to-head per fold, \onenna{}. Resampler minus baseline in $\lvert \mathrm{1NNA} - 0.5 \rvert$; negative favours the resampler. Holm-adjusted p is 0.003 in every fold against VesselVAE and PartVessel. \label{tab:E5}}
\begin{tabular*}{\linewidth}{@{\extracolsep{\fill}}lccccccc@{}}
\toprule
\multirow{2}{*}{Fold} & \multicolumn{3}{c}{VesselGPT} & \multicolumn{2}{c}{VesselVAE} & \multicolumn{2}{c}{PartVessel} \\
\cmidrule(lr){2-4} \cmidrule(lr){5-6} \cmidrule(lr){7-8}
 & $\Delta$ & 95\% CI & Holm p & $\Delta$ & 95\% CI & $\Delta$ & 95\% CI \\
\midrule
0 & +0.006 & {[}$-$0.082, +0.094{]} & 0.87 & $-$0.312 & {[}$-$0.382, $-$0.247{]} & $-$0.238 & {[}$-$0.309, $-$0.173{]}  \\

1 & $-$0.078 & {[}$-$0.156, $-$0.002{]} & 0.045 & $-$0.356 & {[}$-$0.425, $-$0.287{]} & $-$0.269 & {[}$-$0.338, $-$0.202{]}  \\

2 & +0.029 & {[}$-$0.038, +0.094{]} & 0.37 & $-$0.297 & {[}$-$0.363, $-$0.239{]} & $-$0.231 & {[}$-$0.297, $-$0.176{]}  \\

3 & +0.022 & {[}$-$0.053, +0.092{]} & 0.56 & $-$0.264 & {[}$-$0.340, $-$0.198{]} & $-$0.262 & {[}$-$0.334, $-$0.193{]}  \\

4 & $-$0.014 & {[}$-$0.087, +0.061{]} & 0.72 & $-$0.324 & {[}$-$0.389, $-$0.265{]} & $-$0.216 & {[}$-$0.282, $-$0.154{]}  \\
\bottomrule
\end{tabular*}
\end{table}

\begin{table}[tbp]\centering
\caption{\textbf{On the IntrA dependence AUC, the resampler's lead is detected against PartVessel in every fold and against VesselGPT in none.} IntrA head-to-head per fold, dependence AUC. Resampler minus baseline, on all features and on the feature subset; VesselVAE is not reportable. \label{tab:E5b}}
\begin{tabular*}{\linewidth}{@{\extracolsep{\fill}}lcccc@{}}
\toprule
\multirow{2}{*}{Fold} & \multicolumn{2}{c}{VesselGPT} & \multicolumn{2}{c}{PartVessel} \\
\cmidrule(lr){2-3} \cmidrule(lr){4-5}
 & All features & Subset & All features & Subset \\
\midrule
0 & $-$0.096 & $-$0.042 & $-$0.315 & $-$0.224  \\

1 & $-$0.095 & $-$0.045 & $-$0.280 & $-$0.162  \\

2 & $-$0.053 & $-$0.029 & $-$0.280 & $-$0.253  \\

3 & $-$0.100 & $-$0.056 & $-$0.281 & $-$0.208  \\

4 & $-$0.140 & $-$0.040 & $-$0.324 & $-$0.210  \\
\bottomrule
\end{tabular*}
\end{table}

Holm p for \onenna{} against VesselVAE and PartVessel is 0.003 in every fold. Dependence: against PartVessel the criterion holds in every fold (Holm p 0.01); against VesselGPT the resampler is lower in every fold but every difference is below the IntrA noise threshold (Holm p 0.24--0.48).

\begin{table}[tbp]\centering
\caption{\textbf{The whole-tree data-copying test does not flag the resampler (coronary: underfitting); the branch-record version flags it as copying.} Data-copying test $C_T$ (main $k$). Coronary: mean (SD) over 20 repeats. IntrA: mean $\pm$ SD of the 5 fold means. We read $C_T \le -2$ as copying and $C_T \ge +2$ as underfitting. Half-copy control: 52\% (coronary) / 57\% (IntrA) exact training copies, the rest from the independent sampler. \label{tab:E6}}
\begin{tabulary}{\linewidth}{@{}Lcccc@{}}
\toprule
\multirow{2}{*}{Method} & \multicolumn{2}{c}{Whole trees} & \multicolumn{2}{c}{Branch records} \\
\cmidrule(lr){2-3} \cmidrule(lr){4-5}
 & Coronary & IntrA & Coronary & IntrA \\
\midrule
Training copy & $-$7.94 (0.08) & $-$7.38 $\pm$ 0.33 & $-$7.67 (0.07) & $-$7.87 $\pm$ 0.32 \\
Half-copy control & $-$1.10 (0.14) & $-$2.81 $\pm$ 0.15 & $-$2.44 (0.15) & $-$2.75 $\pm$ 0.06 \\
Real validation (calibration) & $-$0.66 (0.02) & undefined in fold 0 & $-$0.83 (0.08) & $-$0.12 $\pm$ 0.99 \\
Resampler & +2.61 (0.20) & +1.39 $\pm$ 0.60 & $-$4.96 (0.09) & $-$5.30 $\pm$ 0.45 \\
Independent sampler & +6.74 (0.13) & +4.00 $\pm$ 0.45 & +4.07 (0.15) & +3.98 $\pm$ 0.38 \\
VesselGPT & +5.26 (0.08) & +2.86 $\pm$ 0.76 & +3.36 (0.15) & +3.97 $\pm$ 0.64 \\
VesselVAE & +6.59 (0.09) & +7.49 $\pm$ 0.82 & +5.37 (0.41) & +6.97 $\pm$ 0.84 \\
PartVessel & +6.79 (0.09) & +4.21 $\pm$ 1.57 & +6.10 (0.25) & +7.84 $\pm$ 0.93 \\
\bottomrule
\end{tabulary}
\end{table}

Verdicts by the decision rule of Appendix~\ref{app:repro} for the resampler: coronary tree ``underpowered''; IntrA tree ``does not detect copying'' (calibration undefined in fold 0); coronary and IntrA branch-record level ``detects copying'' (below the independent sampler in every repeat and every fold). $C_T$ magnitudes are not comparable across levels or datasets, because set sizes differ.

The tree-level result for the resampler is not a clean negative. On coronary data the tree level fails its power check. On IntrA the half-copy control is flagged ($-2.81$), but it copies whole trees exactly, so the check is lenient, and the real-validation calibration check is undefined in one fold. At the branch-record level, branch records clustered within trees likely make the statistic anti-conservative.

\begin{table}[tbp]\centering
\caption{\textbf{On coronary data no learned baseline lies below the curve at any threshold.} $D_M$ with 95\% CI (\onenna{}: 1,000 paired patient-bootstrap replicates; dependence AUC: the 200 stored repeats) and the verdict by the rule of Section~\ref{sec:q3}: above (worse than the curve), below, no diff. (no difference detected), n.d. (not determined). Near-copy rates are in Table~\ref{tab:E1}. At $q = 1$, VesselGPT's \onenna{} comparison and all of VesselVAE's and PartVessel's are flagged as at the resolution floor shared by several curve points, so their interval spans the range of those points (Appendix~\ref{app:repro}); PartVessel at $q = 10$ falls below every curve point in 14--17\% of replicates. No \onenna{} comparison passes the equivalence test at $\pm$0.045. \label{tab:E7}}
\begin{tabular*}{\linewidth}{@{\extracolsep{\fill}}llcccl@{}}
\toprule
Method & Score & q & D & 95\% CI & Verdict \\
\midrule
\multirow{9}{*}{Independent sampler} & \multirow{3}{*}{\onenna{}} & 1 & +0.049 & {[}$-$0.006, +0.103{]} & no diff. \\
& & 5 & +0.058 & {[}+0.002, +0.103{]} & above \\
& & 10 & +0.058 & {[}+0.003, +0.105{]} & above \\
& \multirow{3}{*}{Dep. AUC all} & 1 & +0.146 & {[}+0.092, +0.202{]} & above \\
& & 5 & +0.154 & {[}+0.101, +0.203{]} & above \\
& & 10 & +0.154 & {[}+0.100, +0.202{]} & above \\
& \multirow{3}{*}{Dep. AUC subset} & 1 & +0.161 & {[}+0.100, +0.228{]} & above \\
& & 5 & +0.167 & {[}+0.103, +0.229{]} & above \\
& & 10 & +0.167 & {[}+0.101, +0.232{]} & above \\
\addlinespace
\multirow{9}{*}{VesselGPT} & \multirow{3}{*}{\onenna{}} & 1 & $-$0.068 & {[}$-$0.237, +0.022{]} & n.d. \\
& & 5 & $-$0.040 & {[}$-$0.119, +0.026{]} & no diff. \\
& & 10 & $-$0.030 & {[}$-$0.102, +0.042{]} & no diff. \\
& \multirow{3}{*}{Dep. AUC all} & 1 & $-$0.009 & {[}$-$0.104, +0.065{]} & no diff. \\
& & 5 & +0.020 & {[}$-$0.058, +0.087{]} & no diff. \\
& & 10 & +0.029 & {[}$-$0.044, +0.097{]} & no diff. \\
& \multirow{3}{*}{Dep. AUC subset} & 1 & $-$0.024 & {[}$-$0.140, +0.062{]} & no diff. \\
& & 5 & +0.010 & {[}$-$0.087, +0.087{]} & no diff. \\
& & 10 & +0.017 & {[}$-$0.071, +0.095{]} & no diff. \\
\addlinespace
\multirow{9}{*}{VesselVAE} & \multirow{3}{*}{\onenna{}} & 1 & +0.062 & {[}$-$0.038, +0.120{]} & n.d. \\
& & 5 & +0.034 & {[}$-$0.033, +0.064{]} & no diff. \\
& & 10 & +0.029 & {[}$-$0.029, +0.060{]} & no diff. \\
& \multirow{3}{*}{Dep. AUC all} & 1 & +0.117 & {[}+0.043, +0.160{]} & above \\
& & 5 & +0.099 & {[}+0.050, +0.132{]} & above \\
& & 10 & +0.096 & {[}+0.053, +0.130{]} & above \\
& \multirow{3}{*}{Dep. AUC subset} & 1 & +0.130 & {[}+0.043, +0.193{]} & above \\
& & 5 & +0.114 & {[}+0.057, +0.169{]} & above \\
& & 10 & +0.110 & {[}+0.058, +0.166{]} & above \\
\addlinespace
\multirow{9}{*}{PartVessel} & \multirow{3}{*}{\onenna{}} & 1 & +0.048 & {[}+0.013, +0.113{]} & above \\
& & 5 & +0.065 & {[}+0.018, +0.113{]} & above \\
& & 10 & +0.044 & {[}+0.017, +0.085{]} & n.d. \\
& \multirow{3}{*}{Dep. AUC all} & 1 & +0.086 & {[}+0.041, +0.143{]} & above \\
& & 5 & +0.102 & {[}+0.056, +0.139{]} & above \\
& & 10 & +0.085 & {[}+0.053, +0.129{]} & n.d. \\
& \multirow{3}{*}{Dep. AUC subset} & 1 & +0.074 & {[}+0.005, +0.159{]} & above \\
& & 5 & +0.086 & {[}+0.020, +0.145{]} & above \\
& & 10 & +0.071 & {[}+0.019, +0.141{]} & n.d. \\
\bottomrule
\end{tabular*}
\end{table}

\begin{table}[tbp]\centering
\caption{\textbf{On IntrA no learned baseline lies below the curve in all five folds.} Each fold has its own curve; a direction is stated only if all 5 folds give it, ``no difference detected'' if no fold is n.d., otherwise ``mixed''. Counts are folds out of 5: below, above, no diff. (no difference detected), n.d. (not determined). VesselVAE's dependence AUC is not reportable. No equivalence test passes. \label{tab:E8}}
\begin{tabular*}{\linewidth}{@{\extracolsep{\fill}}llclcccc@{}}
\toprule
\multirow{2}{*}{Method} & \multirow{2}{*}{Score} & \multirow{2}{*}{q} & \multirow{2}{*}{Verdict} & \multicolumn{4}{c}{Folds} \\
\cmidrule(lr){5-8}
 &  &  &  & below & above & no diff. & n.d. \\
\midrule
\multirow{9}{*}{Independent sampler} & \multirow{3}{*}{\onenna{}} & 1 & mixed & 0 & 0 & 2 & 3 \\
& & 5 & no diff. & 0 & 0 & 5 & 0 \\
& & 10 & no diff. & 0 & 0 & 5 & 0 \\
& \multirow{3}{*}{Dep. AUC all} & 1 & mixed & 0 & 1 & 1 & 3 \\
& & 5 & no diff. & 0 & 4 & 1 & 0 \\
& & 10 & above & 0 & 5 & 0 & 0 \\
& \multirow{3}{*}{Dep. AUC subset} & 1 & mixed & 0 & 1 & 1 & 3 \\
& & 5 & above & 0 & 5 & 0 & 0 \\
& & 10 & above & 0 & 5 & 0 & 0 \\
\addlinespace
\multirow{9}{*}{VesselGPT} & \multirow{3}{*}{\onenna{}} & 1 & mixed & 1 & 0 & 0 & 4 \\
& & 5 & no diff. & 2 & 0 & 3 & 0 \\
& & 10 & no diff. & 2 & 0 & 3 & 0 \\
& \multirow{3}{*}{Dep. AUC all} & 1 & n.d. & 0 & 0 & 0 & 5 \\
& & 5 & mixed & 0 & 1 & 3 & 1 \\
& & 10 & no diff. & 0 & 2 & 3 & 0 \\
& \multirow{3}{*}{Dep. AUC subset} & 1 & n.d. & 0 & 0 & 0 & 5 \\
& & 5 & mixed & 0 & 0 & 4 & 1 \\
& & 10 & no diff. & 0 & 0 & 5 & 0 \\
\addlinespace
\multirow{3}{*}{VesselVAE} & \multirow{3}{*}{\onenna{}} & 1 & mixed & 0 & 3 & 0 & 2 \\
& & 5 & mixed & 0 & 3 & 0 & 2 \\
& & 10 & mixed & 0 & 4 & 0 & 1 \\
\addlinespace
\multirow{9}{*}{PartVessel} & \multirow{3}{*}{\onenna{}} & 1 & n.d. & 0 & 0 & 0 & 5 \\
& & 5 & n.d. & 0 & 0 & 0 & 5 \\
& & 10 & mixed & 0 & 1 & 2 & 2 \\
& \multirow{3}{*}{Dep. AUC all} & 1 & above & 0 & 5 & 0 & 0 \\
& & 5 & above & 0 & 5 & 0 & 0 \\
& & 10 & above & 0 & 5 & 0 & 0 \\
& \multirow{3}{*}{Dep. AUC subset} & 1 & mixed & 0 & 1 & 0 & 4 \\
& & 5 & mixed & 0 & 2 & 0 & 3 \\
& & 10 & mixed & 0 & 4 & 0 & 1 \\
\bottomrule
\end{tabular*}
\end{table}

\begin{table}[tbp]\centering
\caption{\textbf{Real validation trees lie below the curve on both coronary scores, and the jittered copy lies below it on the coronary dependence AUC.} Descriptive. Real validation: the validation-split trees (neither training nor test trees; used for model selection, so held out from scoring but not untouched), scored as if they were a generator; the jittered copy is shown at SD 0.5\,mm (coronary) and 0.1 units (IntrA); further coronary levels are in Table~\ref{tab:E10}. (a) Scores (dependence AUC on all features) and measured branch-record near-copy rate (\%); IntrA: fold means. (b) Gap to the curve $D_M$ at $q = 5$, CIs and verdicts as in Tables~\ref{tab:E7} and \ref{tab:E8}; IntrA folds are counted as below/above/no diff./n.d. The IntrA real-validation \onenna{} is size-matched: test trees subsampled to the validation set size and the curve recomputed at that size. \label{tab:E9}}
\begin{tabular*}{\linewidth}{@{\extracolsep{\fill}}lcccccccccc@{}}
\multicolumn{11}{@{}l}{(a) Scores and near-copy rate} \\
\toprule
\multirow{2}{*}{Control} & \multicolumn{5}{c}{Coronary} & \multicolumn{5}{c}{IntrA} \\
\cmidrule(lr){2-6} \cmidrule(lr){7-11}
 & \onenna{} & Dep. AUC & q1 & q5 & q10 & \onenna{} & Dep. AUC & q1 & q5 & q10 \\
\midrule
Jittered copy & 0.887 & 0.581 & 0.28 & 0.83 & 1.5 & 0.831 & 0.616 & 0.07 & 0.17 & 0.59 \\
Real validation & 0.545 & 0.546 & 2.0 & 6.8 & 12.9 & 0.506 & 0.590 & 1.2 & 7.6 & 13.3 \\
\bottomrule
\end{tabular*}

\medskip
\begin{tabular*}{\linewidth}{@{\extracolsep{\fill}}llcclll@{}}
\multicolumn{7}{@{}l}{(b) Gap to the curve at $q = 5$} \\
\toprule
\multirow{2}{*}{Control} & \multirow{2}{*}{Score} & \multicolumn{3}{c}{Coronary} & \multicolumn{2}{c}{IntrA} \\
\cmidrule(lr){3-5} \cmidrule(lr){6-7}
 &  & D & 95\% CI & Verdict & Verdict & Folds \\
\midrule
\multirow{3}{*}{Jittered copy} & \onenna{} & +0.044 & {[}$-$0.031, +0.096{]} & no diff. & n.d. & 0/0/0/5 \\
& Dep. AUC all & $-$0.231 & {[}$-$0.315, $-$0.150{]} & below & mixed & 3/0/1/1 \\
& Dep. AUC subset & $-$0.141 & {[}$-$0.229, $-$0.055{]} & below & mixed & 2/0/1/2 \\
\addlinespace
\multirow{3}{*}{Real validation} & \onenna{} & $-$0.159 & {[}$-$0.238, $-$0.087{]} & below & below & 5/0/0/0 \\
& Dep. AUC all & $-$0.135 & {[}$-$0.221, $-$0.064{]} & below & no diff. & 2/0/3/0 \\
& Dep. AUC subset & $-$0.092 & {[}$-$0.173, $-$0.012{]} & below & no diff. & 0/0/5/0 \\
\bottomrule
\end{tabular*}
\end{table}
\begin{table}[tbp]\centering
\caption{\textbf{On coronary data a copy with 1\,mm jitter still lies below the curve on the dependence AUC, pooling the arteries changes no score detectably, and on scale-free features no difference from VesselGPT is detected.} Descriptive, coronary only. The resampler is the curve's point at $\lambda = 0$ (Table~\ref{tab:E3}). (a) Training copies with Gaussian point noise of the given SD (radii and topology unchanged; 0.5\,mm is the jittered copy of Table~\ref{tab:E9}; the other levels use different seeds), and the resampler at $\lambda = 0$ fitted per artery (as everywhere else) or pooled across arteries without artery labels. Verdicts against the curve at $q = 1/5/10$ as in Table~\ref{tab:E7}: B below, A above, = no difference detected, n.d. not determined (near-copy rate outside the curve's range), B$^{*}$ below by extrapolation beyond the last curve point; one entry when all three agree. Pooled minus per-artery, 95\% paired intervals: \onenna{} $+$0.008 [$-$0.057, $+$0.073], dependence AUC $+$0.004 [$-$0.078, $+$0.084]. (b) \onenna{} on all 11 tree features and on the 3 scale-free ones (branch count, depth, mean tortuosity): the gap from 0.5, each learned baseline's $\lvert \onenna{} - 0.5 \rvert$ minus the resampler's (95\% paired interval; positive: the learned baseline is farther from 0.5, the opposite sign to Table~\ref{tab:2}), matching the head-to-head rule, and the verdict against the scale-free curve at $q = 5$ (11-feature verdicts: Tables~\ref{tab:E7} and~\ref{tab:E9}). The resampler's scale-free \onenna{} is below 0.5, and VesselGPT's point estimate is the closer to 0.5. \label{tab:E10}}
\begin{tabular*}{\linewidth}{@{\extracolsep{\fill}}lcccccc@{}}
\multicolumn{7}{@{}l}{(a) Perturbation levels and pooled resampler} \\
\toprule
\multirow{2}{*}{Set} & \multicolumn{3}{c}{\onenna{}} & \multicolumn{2}{c}{Dep. AUC} & \multirow{2}{*}{Near-copy (\%)} \\
\cmidrule(lr){2-4} \cmidrule(lr){5-6}
 & Value & Verdict & Scale-free & Value & Verdict &  \\
\midrule
Training copy & 0.545 & n.d. & 0.477 & n/a & n/a & 100 / 100 / 100 \\
Copy, SD 0.1\,mm & 0.551 & n.d. & 0.474 & 0.575 & n.d. & 99.4 / 99.6 / 99.7 \\
Copy, SD 0.25\,mm & 0.577 & n.d. & 0.508 & 0.579 & n.d. & 55.0 / 83.9 / 90.8 \\
Copy, SD 0.5\,mm & 0.887 & = & 0.885 & 0.581 & B & 0.28 / 0.83 / 1.5 \\
Copy, SD 1\,mm & 0.998 & A / n.d. / n.d. & 0.997 & 0.588 & B / B$^{*}$ / B$^{*}$ & 0 / 0 / 0 \\
\addlinespace
Resampler, per artery & 0.628 & curve & 0.463 & 0.613 & curve & 34.2 / 51.4 / 59.1 \\
Resampler, pooled & 0.636 & n.d. & 0.485 & 0.618 & n.d. & 39.9 / 57.3 / 64.7 \\
\bottomrule
\end{tabular*}

\medskip
\begin{tabular*}{\linewidth}{@{\extracolsep{\fill}}lccccc@{}}
\multicolumn{6}{@{}l}{(b) \onenna{} on all and on scale-free tree features} \\
\toprule
\multirow{2}{*}{Set} & \multicolumn{2}{c}{11 features} & \multicolumn{3}{c}{Scale-free (3 features)} \\
\cmidrule(lr){2-3} \cmidrule(lr){4-6}
 & \onenna{} & Gap from 0.5 & \onenna{} & Gap from 0.5 & Verdict \\
\midrule
Training copy & 0.545 &  & 0.477 &  & n.d. \\
Copy, SD 0.5\,mm & 0.887 &  & 0.885 &  & A \\
Real validation & 0.545 &  & 0.445 &  & = \\
\addlinespace
Resampler & 0.628 &  & 0.463 &  & curve \\
Independent sampler & 0.832 &  & 0.538 &  & = \\
VesselGPT & 0.776 & $+$0.148 [$+$0.065, $+$0.225] & 0.533 & $-$0.005 [$-$0.067, $+$0.095] & = \\
VesselVAE & 0.966 & $+$0.338 [$+$0.270, $+$0.402] & 0.694 & $+$0.157 [$+$0.068, $+$0.230] & = \\
PartVessel & 0.999 & $+$0.371 [$+$0.313, $+$0.445] & 0.952 & $+$0.415 [$+$0.325, $+$0.482] & A \\
\bottomrule
\end{tabular*}
\end{table}

\begin{table}[tbp]\centering
\caption{\textbf{On IntrA a copy jittered by a quarter of the point spacing lies below the curve on \onenna{} in all five folds.} Descriptive. Training copies with Gaussian point noise of the given SD in coordinate units (point spacing 0.2; radii and topology unchanged; 0.1 is the jittered copy of Table~\ref{tab:E9}; the other levels share one random stream, scaled). Scores and near-copy rates are fold means; verdicts against each fold's curve at $q = 5$ are counted as below / above / no diff. / n.d. over the 5 folds. At $q = 1$ the 0.05 copy is also below on \onenna{} in 5 of 5 folds; at $q = 10$ the 0.2 copy is below on the dependence AUC in 5 of 5 folds, at the near-copy floor. \label{tab:E11}}
\begin{tabular*}{\linewidth}{@{\extracolsep{\fill}}lcccccc@{}}
\toprule
\multirow{2}{*}{Set} & \multicolumn{3}{c}{\onenna{}} & \multicolumn{2}{c}{Dep. AUC} & \multirow{2}{*}{Near-copy (\%)} \\
\cmidrule(lr){2-4} \cmidrule(lr){5-6}
 & Value & Folds & Scale-free & Value & Folds &  \\
\midrule
Training copy & 0.494 & 0/0/0/5 & 0.496 & n/a & n/a & 100 / 100 / 100 \\
Copy, SD 0.02 & 0.492 & 0/0/0/5 & 0.503 & 0.609 & 0/0/0/5 & 99.4 / 99.5 / 99.6 \\
Copy, SD 0.05 & 0.526 & 5/0/0/0 & 0.591 & 0.608 & 0/0/5/0 & 5.4 / 25.3 / 47.8 \\
Copy, SD 0.1 & 0.831 & 0/0/0/5 & 0.827 & 0.616 & 3/0/1/1 & 0.07 / 0.17 / 0.59 \\
Copy, SD 0.2 & 0.984 & 0/4/0/1 & 0.960 & 0.645 & 4/0/0/1 & 0 / 0 / 0 \\
\bottomrule
\end{tabular*}
\end{table}

\begin{table}[tbp]\centering
\caption{\textbf{Smooth displacement keeps copies near-copies up to far larger SDs than jitter does, and the coronary copy displaced by 4\,mm lies below the curve on the dependence AUC.} Descriptive. Training copies whose points are moved by a smooth Gaussian displacement field (correlation length 5 point spacings; radii and topology unchanged) of the given SD, in mm (coronary, spacing 1\,mm) or coordinate units (IntrA, spacing 0.2); levels share one random stream, scaled; the two largest levels per dataset are set by a rule that reads validation data only. Coronary: 3 seeds of 200 trees, verdict against the curve at $q = 5$ (B below, A above, = no difference detected, n.d. not determined). IntrA: fold means, verdicts counted over the 5 folds as below / above / no diff. / n.d. at $q = 5$. Near-copy rate at $q = 5$. Point re-sampling of the training copy and of the test trees is summarised in Appendix~\ref{app:disclosures}. \label{tab:E12}}
\begin{tabular*}{\linewidth}{@{\extracolsep{\fill}}lcccccc@{}}
\toprule
\multirow{2}{*}{Set} & \multicolumn{2}{c}{\onenna{}} & \multicolumn{2}{c}{Dep. AUC} & \multirow{2}{*}{Scale-free \onenna{}} & \multirow{2}{*}{Near-copy (\%)} \\
\cmidrule(lr){2-3} \cmidrule(lr){4-5}
 & Value & Verdict & Value & Verdict &  &  \\
\midrule
\multicolumn{7}{@{}l}{\emph{Coronary}} \\
Smooth, SD 0.1\,mm & 0.543 & n.d. & 0.574 & n.d. & 0.468 & 99.97 \\
Smooth, SD 0.25\,mm & 0.543 & n.d. & 0.575 & n.d. & 0.467 & 99.72 \\
Smooth, SD 0.5\,mm & 0.546 & n.d. & 0.581 & n.d. & 0.487 & 98.82 \\
Smooth, SD 1\,mm & 0.552 & n.d. & 0.573 & n.d. & 0.496 & 89.46 \\
Smooth, SD 2\,mm & 0.668 & = & 0.578 & = & 0.634 & 18.86 \\
Smooth, SD 4\,mm & 0.976 & A & 0.592 & B & 0.960 & 0.83 \\
\addlinespace
\multicolumn{7}{@{}l}{\emph{IntrA}} \\
Smooth, SD 0.02 & 0.493 & 0/0/0/5 & 0.608 & 0/0/0/5 & 0.491 & 99.67 \\
Smooth, SD 0.05 & 0.492 & 0/0/0/5 & 0.608 & 0/0/0/5 & 0.493 & 98.99 \\
Smooth, SD 0.1 & 0.491 & 0/0/0/5 & 0.609 & 0/0/0/5 & 0.489 & 97.06 \\
Smooth, SD 0.2 & 0.505 & 0/0/0/5 & 0.608 & 0/0/0/5 & 0.533 & 69.24 \\
Smooth, SD 0.4 & 0.634 & 4/0/1/0 & 0.610 & 1/0/4/0 & 0.695 & 2.16 \\
Smooth, SD 0.8 & 0.913 & 0/0/0/5 & 0.631 & 4/0/0/1 & 0.877 & 0.02 \\
\bottomrule
\end{tabular*}
\end{table}

\FloatBarrier

\section{Meshing (coronary, n = 20 per method)}\label{app:meshing}

Each tree was converted to a surface and volume mesh with a fixed tool; success was judged by a fixed criterion. Real test 11/20, resampler 14/20, VesselGPT 2/20. Fisher exact test, resampler against VesselGPT: p = 2.4 $\times$ 10$^{-4}$. With n = 20, below its planned size, this shows only that resampled trees can be meshed; no validity ranking is claimed.

\FloatBarrier

\section{Galleries}\label{app:galleries}

\begin{figure}[tbp]\centering
\includegraphics[width=\linewidth]{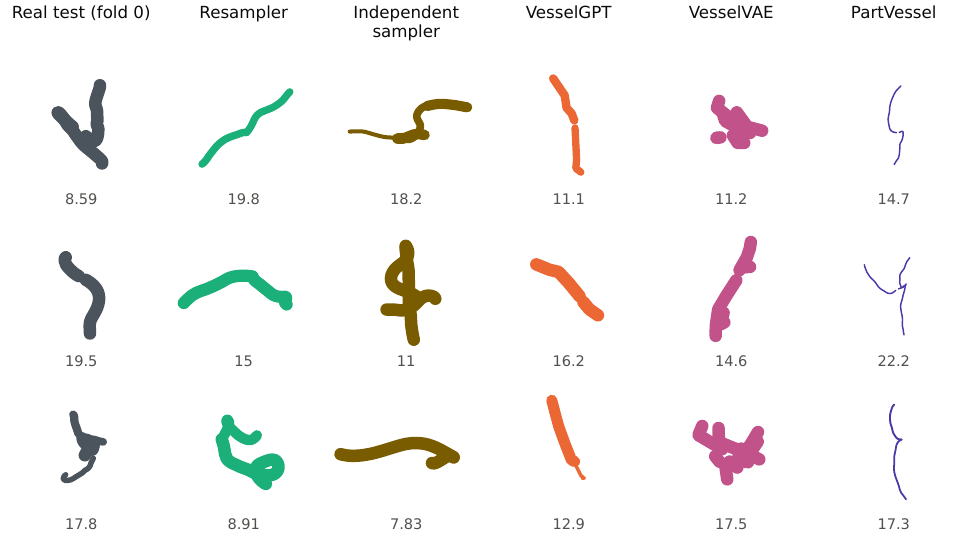}
\caption{\textbf{In random IntrA draws, VesselVAE's trees are tangled and PartVessel's thin, unlike real ones.} Same layout and drawing procedure as Figure~\ref{fig:4} (random draws with a fixed seed, no selection, same renderer), fold 0; the number under each tree is its largest extent in the coordinate units of the IntrA centerlines. \label{fig:G1}}
\end{figure}

\clearpage
\section{Specification of the evaluation and the resampler}\label{app:spec}

This appendix specifies every quantity behind the reported numbers: the features (Appendix~\ref{app:spec-features}), the resampler (Appendix~\ref{app:spec-resampler}), the inference procedures (Appendix~\ref{app:spec-inference}) and the data splits (Appendix~\ref{app:spec-data}). It was written from the code that produced the results. Lengths are in mm (for IntrA, the coordinate unit of its centerlines), angles in degrees unless stated.

\subsection{Tree features and branch records}\label{app:spec-features}

\textbf{Input and extraction conventions.} A tree is a set of branches, each an ordered polyline of points $(x, y, z, r)$, with a list of parent-child edges. A tree that cannot be parsed, has no branches, contains a non-finite value or a value above $10^4$ in absolute size, or whose features are non-finite, is a \emph{failure tree}: its 11 tree features are all 0 and it has no branch records. Missing trees of a generated set of nominal size count as failure trees; a file whose content duplicates another in the same set is dropped and, like a missing tree, replaced by a failure tree. Parsed trees then go through three steps.
\begin{enumerate}
\item \emph{Spanning tree.} Edges to unknown branches, self-edges and duplicate edges are dropped, and the tree is traversed breadth first from its roots (branches without a parent, branch ``0'' first); branches not reached become further roots, and an edge into an already visited branch is dropped.
\item \emph{Topology normalisation}, repeated until nothing changes: (i) a non-root branch of zero length that has children is contracted, its children attaching to its parent (this can create junctions with three or more children); (ii) a branch with exactly one child is merged with it into one polyline (the child's first point is dropped if it lies within $10^{-6}$ of the parent's last point). Junctions with three or more children are kept as they are.
\item \emph{Resampling.} Each branch polyline, after removing repeated consecutive points, is resampled at uniform arc length $\Delta$ (coronary 1.0 mm, IntrA 0.2) by linear interpolation of $x, y, z, r$; the end point is kept.
\end{enumerate}
Depth is 0 for the root and parent depth plus 1 otherwise. For branch $b$: $L_b$ is the arc length of its resampled polyline, $c_b$ the distance between its first and last points, $\bar r_b$ the mean radius of its points and $s_b$ its taper, the least-squares slope of radius against arc length (mm per mm).

\begin{table}[htbp]\centering
\caption{\textbf{\onenna{} compares trees through 11 summaries of size, branching, length, tortuosity, radius and extent.} The 11 tree features (used by \onenna{} and the tree-level near-copy rate), computed after normalisation and resampling. $\langle y \rangle_L = \sum_b L_b y_b / \sum_b L_b$ is the length-weighted mean over branches. \label{tab:H1}}
\begin{tabular*}{\linewidth}{@{\extracolsep{\fill}}rlp{0.56\linewidth}l@{}}
\toprule
 & Feature & Definition & Unit \\
\midrule
1 & Branch count & number of branches & count \\
2 & Maximum depth & largest branch depth (root 0) & count \\
3 & Total length & $\sum_b L_b$ & mm \\
4 & Mean branch length & $\langle L \rangle_L = \sum_b L_b^2 / \sum_b L_b$ (length-weighted) & mm \\
5 & SD of branch length & $\big(\langle (L - \langle L \rangle_L)^2 \rangle_L\big)^{1/2}$ (length-weighted) & mm \\
6 & Mean tortuosity & $\langle L_b / c_b \rangle_L$ over branches with $L_b > 0$ and $c_b > 10^{-9}$; 1 if there is none & 1 \\
7 & Mean radius & $\langle \bar r \rangle_L$ (unweighted mean of $\bar r_b$ if the total length is 0) & mm \\
8 & Mean taper & $\langle s \rangle_L$ over branches with at least 2 points and $L_b > 0$; 0 if there is none & mm/mm \\
9--11 & Extents & square roots of the eigenvalues, largest first, of the covariance matrix (divisor $n$) of the $(x, y, z)$ of all resampled points & mm \\
\bottomrule
\end{tabular*}
\end{table}

\begin{table}[htbp]\centering
\caption{\textbf{A branch record describes a branch together with its parent and closest sibling in 10 numbers.} The 10 coordinates of a branch record, one record per non-root branch $b$ with parent $p$ and closest sibling $q$. A window radius is the median radius over an arc-length window measured from the branch's start (or end): it skips $a = \min(D_b, L_b/3)$ and ends at $\min(a + 2D_b, L_b)$, with $D_b = \max(2 \times \text{median radius of } b,\ \Delta)$; an empty window uses the point nearest its centre. The dependence subset is coordinates 2, 3, 6, 7, 9 and 10. \label{tab:H2}}
\begin{tabular*}{\linewidth}{@{\extracolsep{\fill}}rlp{0.56\linewidth}l@{}}
\toprule
 & Coordinate & Definition & Unit \\
\midrule
1 & $d_b$ & depth of $b$ (at least 1) & count \\
2 & $L_b$ & arc length & mm \\
3 & $r^{\mathrm{start}}_b$ & window radius at the start & mm \\
4 & $r^{\mathrm{end}}_b$ & window radius at the end & mm \\
5 & $s_b$ & taper; missing if fewer than 2 points or $L_b = 0$ & mm/mm \\
6 & $\tau_b$ & $L_b / c_b$; missing if $L_b = 0$ or $c_b \le 10^{-9}$ & 1 \\
7 & $L_p$ & arc length of the parent & mm \\
8 & $r^{\mathrm{end}}_p$ & window radius at the parent's end & mm \\
9 & $L_q$ & arc length of the closest sibling & mm \\
10 & $r^{\mathrm{start}}_q$ & window radius at the closest sibling's start & mm \\
\bottomrule
\end{tabular*}
\end{table}

\textbf{Closest sibling and non-binary junctions.} At a binary junction $q$ is the other child. At a junction with three or more children, every child gets its own branch record, and $q$ is the sibling whose start direction makes the smallest angle with that of $b$. A start direction is the principal axis (least-squares line) of the points within arc length 2.5 mm (coronary) or 1.0 (IntrA) of the branch start, oriented away from it (the first segment if fewer than 2 points fall in that range); a zero-length branch has no direction and counts as an infinite angle; ties go to the first sibling in stored order. After normalisation every non-root branch has a sibling.

\textbf{Standardisation and missing values.} \onenna{}: tree features are z-scored with the mean and SD (divisor $n - 1$) of the real training trees (coronary: both arteries pooled; IntrA: per fold); an SD below $10^{-6}$ times the absolute mean is replaced by 1. Tree features are never missing. Near-copy rate: branch records are z-scored with the mean and SD (divisor $n$, plus $10^{-9}$) of the training branch records over finite values, and a missing value is then set to 0 (the training mean); distances are Euclidean. Dependence test: missing values stay missing, are not ranked and go to the classifier as missing.

\subsection{The resampler}\label{app:spec-resampler}

\textbf{Fitting.} The resampler reads the raw training trees (no normalisation or resampling). Per branch: $L$ is the arc length of its polyline; $\tau = L / c$, or 1 if $L = 0$ or $c \le 10^{-6}$; a least-squares line of radius against arc length (at least 3 points and positive length; otherwise the first and last radius and slope 0) gives the start radius $r^{\mathrm{start}}$ (intercept), the end radius $r^{\mathrm{end}}$ (value at $L$) and the slope $s$. End and start tangents are the chords over the last or first $w_t$ of a branch. Every branch with at least two children is a training bifurcation (with three or more children, a fixed random pair is used; a single child counts as no split). Its key is $(d, \log \max(L, 10^{-3}), \log \max(r^{\mathrm{end}}, 10^{-3}), \tau)$, standardised by the mean and SD (divisor $n$) of all training keys. Its payload holds, per daughter, $L$, $\tau$, $s$, $\rho = r^{\mathrm{start}}_{\mathrm{daughter}} / \max(r^{\mathrm{end}}_{\mathrm{parent}}, 10^{-6})$, the angle $\alpha$ between the parent's end tangent and the daughter's start tangent, and the split flag $\beta$ (the daughter has at least two children); plus $\Delta\phi$, the azimuth of the second daughter minus that of the first around the parent's end tangent, modulo $2\pi$. Undefined angles (no tangent) are replaced by the pooled median. Each training root gives a root record $(L, \tau, s, r^{\mathrm{start}}, \beta)$. Radii are clipped to $[r_{\min}, r_{\max}]$, the smallest positive and the largest training radius.

\textbf{Caps.} The branch cap is the smaller of 20 and the largest branch count of a training tree, the depth cap the smaller of 8 and the largest training depth, and a branch has at most 400 points. Fitted values are in Table~\ref{tab:H3}. A tree whose branch count exceeds the branch cap is discarded and drawn again (at most 200 attempts); a branch at the depth cap does not split, so the depth cap truncates rather than rejects.

\textbf{Generation} (one model per artery on coronary data, one per fold on IntrA; $\lambda = 0$ gives the resampler without noise):
\begin{enumerate}
\item Draw a training root record uniformly and perturb it (Appendix~\ref{app:repro}). The root starts at the origin, in a direction uniform on the sphere, with radius $r^{\mathrm{start}}$.
\item Take open branches first in, first out. For a branch with target length $L$, tortuosity $\tau$, slope $s$, start radius $r_0$, start point and direction, draw its path (below) with $n = \min(400, \mathrm{round}(L/\Delta) + 1)$ points and radii $r_i = \mathrm{clip}(r_0 + s\Delta i,\, r_{\min}, r_{\max})$, $i = 0, \dots, n-1$.
\item If its split flag is 0 or its depth equals the depth cap, it stays a leaf. Otherwise form its key from the target $L$ and $\tau$ and the drawn end radius $r_{n-1}$. Take the $K = 10$ training bifurcations at the same depth with the nearest keys (Euclidean on the three standardised continuous coordinates); if that depth has fewer than $K$, take the $K$ nearest over all depths with standardised depth as a fourth coordinate. Pick one uniformly and perturb its payload.
\item Attach the two daughters in random order. The first gets azimuth $\varphi_0 \sim U(0, 2\pi)$, the second $\varphi_0 + \Delta\phi$; a daughter's start direction is the parent's end tangent on the drawn path rotated by $\alpha$ towards its azimuth; it starts at the parent's last point with radius $\mathrm{clip}(r_{n-1}\rho, r_{\min}, r_{\max})$ and copies $L$, $\tau$, $s$ and $\beta$.
\item If the tree now has more branches than the branch cap, discard it and restart at step 1.
\end{enumerate}
Noise draws come from a random stream separate from the sampling stream, so $\lambda = 0$ reproduces the noise-free sets exactly. A value with $\tau \le 1 + 10^{-9}$ is never perturbed and is left out of $\sigma_{\log(\tau - 1)}$, as $\log(\tau - 1)$ is undefined there; in the training data these are branches of zero length (whose length also stays 0) and exactly straight branches. Likewise a non-positive $\rho$ or inlet radius is left unchanged. Noise SDs are computed per model, daughter values pooled over both daughters of all training bifurcations and root values over training roots.

\textbf{Path generation and calibration table.} A path is a smooth random walk of $n$ points with step $\Delta$: the heading $h$ starts at the given direction; each step adds $\Delta h$ to the position, then updates a curvature vector $w \leftarrow \varrho w + (1 - \varrho^2)^{1/2} \sigma \xi$ with $\xi \sim \mathcal N(0, I_3)$ and $\varrho = e^{-\Delta/\ell}$, projects $w$ perpendicular to $h$, and sets $h \leftarrow (h + \Delta w)/\lVert h + \Delta w \rVert$ ($w$ starts as $\sigma$ times a standard normal vector in the plane perpendicular to $h$). For each branch, 32 candidate paths are drawn and the one whose tortuosity is closest to the target $\tau$ is kept. The curvature scale $\sigma$ comes from a calibration table built once per run (per dataset; per fold on IntrA) with random seed 12345: rows are the distinct values of 24 log-spaced point counts from 2 to $n_{\max} = \min(400, \lfloor L_{\max}/\Delta \rfloor + 1)$ ($L_{\max}$ the longest training root or daughter of the run), columns are 40 log-spaced $\sigma$ from $10^{-3}$ to 3 radians per unit length, and each cell is the median tortuosity of 64 simulated paths, made non-decreasing along each row by a running maximum. A target is inverted in the row with the nearest point count: $\log \sigma$ is interpolated linearly in $\tau$, and targets below or above the row's range get the smallest or largest $\sigma$.

\begin{table}[htbp]\centering
\caption{\textbf{Each dataset has one set of constants, shared by every method.} Constants of the evaluation and the resampler. Caps are the fitted values. \label{tab:H3}}
\begin{tabular*}{\linewidth}{@{\extracolsep{\fill}}lll@{}}
\toprule
Constant & Coronary & IntrA \\
\midrule
Point spacing $\Delta$ (evaluation resampling and path step) & 1.0 mm & 0.2 \\
Direction window for the closest sibling & 2.5 mm & 1.0 \\
Resampler tangent window $w_t$ & 3.0 mm & 1.0 \\
Path correlation length $\ell$ & 5.0 mm & 2.0 \\
Neighbours $K$ & 10 & 10 \\
Branch cap & RCA 15, LCA 19 & 9 \\
Depth cap & RCA 6, LCA 7 & 4 (fold 3: 3) \\
Jittered copy: SD of point noise & 0.5 mm & 0.1 \\
\bottomrule
\end{tabular*}
\end{table}

\textbf{Validity.} A generated tree is valid if it parses, is non-empty, every edge joins two existing branches, every branch has at most one parent, there is a single root, the tree is connected and acyclic, every branch has at least 2 points, and every value is finite and at most $10^4$ in absolute size. Invalid trees that parse are still scored; only failure trees (above) receive the all-zero features. In the resampler's coronary validation runs the only check that fails is the two-point rule, from branches whose target length is below about $\Delta/2$. Share of valid trees on coronary validation runs (1{,}000 trees per $\lambda$): 0.98 at $\lambda = 0$, between 0.96 and 0.97 up to $\lambda = 1$, 0.92 at $\lambda = 1.5$, 0.87 at $\lambda = 2.0$ and 0.68 at $\lambda = 3.0$, against 0.97 for real coronary training and test trees. The flag does not count branch collisions; their share rises from 0.11 at $\lambda = 0$ to 0.42 at $\lambda = 3.0$. Cap rejection stays below 1\% of attempts at every $\lambda$ (at most 0.7\%). On IntrA the flag is uninformative: real IntrA test trees also score only 0.17 to 0.28 across folds, because single-point branches occur in the data, so no IntrA validity trend is claimed.

\subsection{Inference procedures}\label{app:spec-inference}

Every interval below is conditional on the stored generated sets: no generator is retrained or regenerated, and only the stored seed sets of a method enter (coronary 3 per method; IntrA 1 or 2 per fold). The sampling unit for real data is always the patient.

\textbf{\onenna{} and its patient bootstrap.} For a generated tree set $G$ and real trees $R$ (z-scored tree features), a tree is classified correctly if its nearest neighbour in its own set is strictly closer than its nearest neighbour in the other set (a tie counts as wrong); \onenna{} is the share classified correctly.
\begin{enumerate}
\item Point estimate: for each stored seed set, the mean over 20 subsamples of $|R|$ trees drawn without replacement, scored against all test trees; then the mean over seed sets.
\item Replicate $b = 1, \dots, 1000$: draw the test patients with replacement and take all their trees ($R^*_b$); draw $S$ seed sets with replacement from the $S$ stored ones; from each drawn set take $|R^*_b|$ trees without replacement (failure trees fill a set that is too small); compute \onenna{} for each and average.
\item Duplicates: a patient drawn $m$ times contributes $m$ copies of each of its trees; copies of the same tree are excluded from each other's candidate neighbours, so a copy is never its own nearest neighbour; other trees of the same patient remain candidates. Generated sets are never drawn with replacement within a replicate.
\item The 95\% interval is the 2.5 and 97.5 percentiles of the replicates. A paired comparison of methods A and B uses the same $R^*_b$ and the same seed-draw indices for both, and reports $\Delta\lvert \onenna{} - 0.5 \rvert$ per replicate, with two-sided $p = 2\min(f, 1 - f)$, floored at $1/B$, where $f$ is the share of replicates in which A is better (ties count one half).
\end{enumerate}

\textbf{One dependence repeat} ($r = 0, \dots, 199$; the same steps on validation data when calibrating).
\begin{enumerate}
\item Real class: draw the test patients with replacement and take all their trees. The random stream depends only on $r$, so every method in repeat $r$ faces the same real resample.
\item Generated class: as many trees as the real resample holds, drawn without replacement from the stored sets, split evenly over seed sets (remainder to randomly chosen sets) and, when the trees carry artery labels, with the real resample's numbers of left and right trees. The training copy here is the real training set drawn in the same way.
\item Take all branch records of the drawn trees (failure trees have none) and weight each by $1/(\text{branch records of its tree})$, in fitting and in the AUC. Because classifier fitting and AUC use equal-tree weights, the score can also reflect associations between feature values and tree size.
\item Groups: real patient (all copies of a resampled patient form one group); generated tree; for the training and jittered copies, the source training patient.
\item Cross-validation: scikit-learn's \texttt{GroupKFold} with 5 folds and no shuffling, which has no random component (largest groups first, each to the fold with fewest samples); partitions differ between repeats only because the resampled data differ.
\item Empirical-CDF transform, per cross-validation fold, class and column, fitted on that class's branch records in the training folds only: a training-fold value becomes its rank among them, ties broken uniformly at random, divided by $n + 1$; a held-out value $x$ becomes $(a + V(c - a + 1))/(n + 1)$, with $a$ and $c$ the numbers of training-fold values below and at most $x$ and $V \sim U(0, 1)$. A tie block is thus spread uniformly over its rank range.
\item Classifier: \texttt{HistGradientBoostingClassifier} with 200 iterations and otherwise default settings (learning rate 0.1, 31 leaves, at least 20 samples per leaf, no L2 penalty, 255 bins, native handling of missing values), random state $r$, sample weights as in step 3. Its default early stopping switches on only above 10,000 training samples; every fit here is far smaller (on coronary data about 1,100 branch records per repeat in both classes together), so all 200 iterations run.
\item AUC: weighted ROC AUC of the out-of-fold probabilities, pooled over the folds on coronary data and the mean of the 5 per-fold AUCs on IntrA (few real patients make IntrA folds unbalanced). Shuffle sanity repeats steps 6--8 after permuting every column independently within each class.
\end{enumerate}
A method's dependence AUC is the mean over the 200 repeats, its interval the 2.5 and 97.5 percentiles; paired differences are taken repeat by repeat. The spread over repeats combines test-patient resampling, generated-tree subsampling, partitions and tie draws. The calibration gate compares the column-permuted training copy (positive control) with the training copy (null): it passes if the positive control's 2.5 percentile exceeds the null's 97.5 percentile on both feature sets.

\textbf{Noise thresholds.} On validation data (real class: validation patients), with 200 repeats, take pairs of generators that should be indistinguishable: on coronary data the three jittered-copy seed sets in pairs and the training copy against the jittered copy; on IntrA the training copy against the jittered copy, in fold 0 only (7 validation patients), with the fold-mean AUC. For each pair, compute the SD of the paired per-repeat differences; the threshold is twice the largest SD over pairs, per feature set (coronary SDs 0.0471 and 0.0492; IntrA 0.0888 and 0.0778), giving the values in Appendix~\ref{app:dependence}.

\textbf{Gap to the curve $D_M$} (per dataset; on IntrA per fold, each fold with its own curve and bootstrap).
\begin{enumerate}
\item Fixed inputs: the generated sets of the curve's 9 noise levels $\lambda \in \{0, 0.1, 0.25, 0.5, 0.75, 1.0, 1.5, 2.0, 3.0\}$ and of the compared method; the 200 stored dependence repeats of each. The curve's values are known only at these levels.
\item Floor: $\phi = 0.5/N_{\min}$, with $N_{\min}$ the smallest branch-record count among the curve levels and the compared methods; every rate becomes $\max(\text{rate}, \phi)$ and is used on a $\log_{10}$ scale.
\item Curve value $\hat f(c)$ (the same rule in every replicate): if $c = \phi$ and at least two levels sit at $\phi$, the range of their values (tie); otherwise linear interpolation between the first $\lambda$-consecutive pair of levels whose rates bracket $c$ (the mean if the pair has equal rates); otherwise the value of the level nearest in rate (copies less than every level: extrapolated low; more: extrapolated high).
\item $D_a = f_M - \max \hat f$, $D_b = f_M - \min \hat f$ (equal unless tie); $D_M$ is their midpoint. The point estimate uses the full-reference \onenna{} or the 200-repeat mean dependence AUC, and full-set rates.
\item \onenna{} replicates ($B = 1000$): one patient resample $R^*_b$ shared by the method and every level; the thresholds $t_q(b)$ are the $q$-th percentiles of the nearest-training distances of the real test branch records of $R^*_b$, counted with multiplicity; each set's \onenna{} as above, and its near-copy rate from the branch records of a tree bootstrap (with replacement, full size) of each drawn seed set, pooled.
\item Dependence replicates: the 200 stored repeats; thresholds from repeat $r$'s real resample; near-copy rate from a tree bootstrap within every stored seed set, drawn independently of the trees used for that repeat's AUC; repeats with a non-finite value are dropped.
\item 95\% interval: from the 2.5 percentile of $D_a$ to the 97.5 percentile of $D_b$ (90\%: 5 and 95). A status (tie, extrapolated) flags the comparison if it holds at the point estimate or in more than 5\% of replicates; verdicts and the IntrA 5/5 rule are in Appendix~\ref{app:repro} and Table~\ref{tab:E8}.
\end{enumerate}
The interval therefore covers test-patient resampling, the resampled near-copy threshold and tree resampling within the stored sets. It holds fixed the stored generated sets and their seeds (for \onenna{}, seed sets are resampled among themselves), the trained models and the noise levels; it excludes interpolation error between levels; for a method with no near-copies, only the threshold moves its rate, so the floor and the tie rule, not the interval, express how little such a rate is known.

\textbf{Permutation null for two real samples} (coronary; whole patients are permuted).
\begin{enumerate}
\item Pool the 496 training and test patients (446 and 50), both arteries of a patient together.
\item In each of 200 permutations, assign 50 randomly chosen patients to the test role and the rest to the training role; run the training-copy comparison (training-role trees as the generated class) with 40 dependence repeats; record the mean AUC.
\item Observed value: the true split, also with 40 repeats (0.574; the 200-repeat value is 0.577). One-sided $p = (1 + \#\{\text{null} \ge \text{observed}\})/201 = 0.04$.
\end{enumerate}
The null has mean 0.548 (95\% range 0.523--0.578). Two samples of exchangeable patients therefore score about 0.55, not 0.5: most of the training copy's excess over 0.5 comes from the pipeline. The true split lies at the upper edge of the null; this is either a small genuine difference between the training and test patients (about 0.03 in AUC) or chance at the 4\% level. The level of two real samples is therefore the training copy's dependence AUC, not 0.5. On the six-column subset the observed 0.546 is within the null (mean 0.532, $p = 0.12$).

\subsection{Data}\label{app:spec-data}

\textbf{Coronary extraction.} The coronary trees were extracted in-house from the segmentations of the public ImageCAS dataset \citep{zeng2023imagecas}: one left and one right centerline tree per patient, each branch a polyline of points $(x, y, z, r)$ in mm in patient coordinates at a median point spacing of 1.01 mm, with the root's first point as the inlet. Patients (both arteries together) were split into 446 / 50 / 50 for training, validation and test (892 / 100 / 100 trees), and the trees were pruned as described in Appendix~\ref{app:repro}. The derived trees are not redistributed.

\textbf{IntrA folds.} Pool: 785 trees from 72 patients (patient = IntrA case number). Patients are sorted by decreasing tree count (ties in random order) and assigned one by one to the fold with the fewest trees so far; fold $k$ is the test set of split $k$, 7 validation patients are drawn at random from the other folds, and the remaining patients form the training set (random seed 0 throughout). Table~\ref{tab:H4} lists the patients.

\begin{table}[htbp]\centering
\caption{\textbf{IntrA folds split patients, so no patient's trees are in both training and test.} IntrA splits by patient (IntrA case numbers). Training patients are all patients not listed for that fold. Trees: training / validation / test. \label{tab:H4}}
\begin{tabular*}{\linewidth}{@{\extracolsep{\fill}}lp{0.38\linewidth}p{0.31\linewidth}l@{}}
\toprule
Fold & Test patients & Validation patients & Trees \\
\midrule
0 & 2, 3, 54, 129, 136, 138, 142, 159, 160, 161, 162, 164, 181, 185, 188 & 9, 31, 119, 121, 134, 137, 178 & 543 / 83 / 159 \\
1 & 19, 23, 34, 44, 58, 116, 140, 144, 148, 155, 163, 171, 180, 183 & 1, 134, 135, 139, 166, 170, 175 & 557 / 72 / 156 \\
2 & 1, 28, 31, 40, 42, 119, 128, 139, 149, 152, 153, 165, 177, 187 & 32, 55, 162, 163, 164, 180, 182 & 561 / 68 / 156 \\
3 & 9, 11, 25, 26, 27, 32, 55, 125, 134, 137, 166, 167, 168, 173, 174 & 1, 19, 28, 172, 177, 180, 187 & 555 / 72 / 158 \\
4 & 6, 117, 120, 121, 135, 151, 157, 158, 170, 172, 175, 178, 182, 186 & 32, 55, 58, 116, 137, 166, 168 & 547 / 82 / 156 \\
\bottomrule
\end{tabular*}
\end{table}

With these definitions, constants and splits, the evaluation and the resampler can be reimplemented without our code; the coronary trees themselves are not distributed.

\end{document}